\documentclass[letter,scriptaddress,twocolumn, prl,showkeys,showpacs,groupedaddress]{revtex4-1}

	\usepackage{amsmath}
	\usepackage{morefloats}
	\usepackage{makeidx}
	\usepackage{amsfonts}
	\usepackage{graphicx, xcolor}  
	\usepackage{dcolumn}   
	\usepackage{bm}        
	\usepackage{amssymb}   
	\usepackage[ansinew]{inputenc}
	\usepackage[usenames,dvipsnames]{pstricks}
	\usepackage{epstopdf}
	\usepackage{subfig}
	\usepackage{caption}
	\usepackage{pst-grad} 
	\usepackage{pst-plot} 
	\usepackage[colorlinks,hyperindex]{hyperref}
	\hypersetup
	{
		colorlinks,%
		citecolor=black,%
		linkcolor=black,%
		urlcolor=black,%
	}
	\usepackage{auto-pst-pdf}
	\usepackage{setspace}
	\usepackage{verbatim}

\newcommand{\be}{\begin{equation}}
\newcommand{\ee}{\end{equation}}
\newcommand{\ba}{\begin{array}}
\newcommand{\ea}{\end{array}}

\newcommand{\tr}{\mathop{\mathrm{tr}}}

\newcommand{\mh}{\mathcal{H}}
\newcommand{\mf}{\mathcal{F}}
\newcommand{\mg}{\mathcal{G}}

\newcommand{\ms}{\mathcal{S}}

\newcommand{\mi}{\mathcal{I}}

\newcommand{\mpp}{\mathcal{P}}
\newcommand{\mm}{\mathcal{M}}

\newcommand{\mz}{\mathcal{Z}}

\newcommand{\rd}{{\rm d}}

\newcommand{\rw}{\rm w}

\begin{document}

\title{Adaptive Elastic Networks as Models of Supercooled Liquids}
\author{Le Yan and Matthieu Wyart}
\affiliation{Center for Soft Matter Research, Department of Physics, New York University\\ 4 Washington Place, New York, 10003, NY, USA}

\begin{abstract}
The thermodynamics and dynamics of supercooled liquids correlate with their elasticity. In particular for covalent networks,   the jump of specific heat is  small and the liquid is {\it strong} near the threshold valence where the network acquires rigidity. By contrast, the jump of specific heat and the fragility are large away from this threshold valence. In a previous work [Proc. Natl. Acad. Sci. U.S.A., 110, 6307 (2013)], we could explain these behaviors by introducing a  model of supercooled liquids in which local rearrangements interact via elasticity. However, in that model the disorder characterizing elasticity was frozen, whereas it is itself a dynamic variable in supercooled liquids. Here we study numerically and theoretically adaptive elastic network models where polydisperse springs can move on a lattice, thus allowing for the geometry of the elastic network to fluctuate and evolve with temperature. We show numerically that our previous results on the relationship between structure and thermodynamics hold in these models. We introduce an approximation where redundant constraints (highly coordinated regions where the frustration is large) are treated as an ideal gas, leading to analytical predictions that are accurate in the range of parameters relevant for real materials. Overall, these results lead to a description of supercooled liquids, in which the distance to the rigidity transition controls the number of directions in phase space that cost energy and the specific heat.

\end{abstract}

\maketitle

\section{I. Introduction}

Liquids undergo a glass transition toward an amorphous  solid state when cooled rapidly enough to avoid  crystallization ~\cite{Debenedetti01}. The glass lacks  structural order: it is a liquid ``frozen'' in a local minimum in the energy landscape, due to the slowing down of relaxation processes. It is very plausible that the thermodynamics and the dynamics in supercooled liquids strongly depend on the microscopic structure of these configurations -- hereafter referred to as ``inherent structures"~\cite{Stillinger84}.  However, a majority of glass theories~\cite{Adam65,Kirkpatrick89,Lubchenko07,Bouchaud04,Chamberlin99,Dyre06,Chandler10} have focused on explaining the correlations between macroscopic observables seen in experiments (such as the relationship between thermodynamics and dynamics ~\cite{Martinez01,Wang06}), while only a few~\cite{Hall03,Bevzenko09,Souza09,Rabochiy13} have investigated  the role of structure. 

Experiments reveal that elasticity plays a key role in both the thermodynamic and dynamical properties in supercooled liquids, such as the jump of specific heat and the fragility characterizing the glass transition. Specifically, it has been found that 
(I) glasses present an excess of low-frequency vibrational modes with respect to Debye modes. The number of these excess anomalous modes, quantified as the intensity of the boson peak~\cite{Phillips81}, shows a strong anti-correlation with the fragility~\cite{Ngai97,Novikov05}. (II) The rigidity of the inherent structures is tunable by changing the fraction of components with different valences in network glasses~\cite{Tatsumisago90,Kamitakahara91,Selvanathan99}, where atoms interact via covalent bonds and much weaker Van der Waals force. 
The covalent network becomes rigid~\cite{Maxwell64,Phillips79,Phillips85}, when the average valence $r$ exceeds a threshold $r_c$, determined by the balance between the number of covalent constraints and the degrees of freedom of the system. 
Both the fragility and the jump of specific heat depend non-monotonically on $r$, and their minima coincide with $r_c$~\cite{Tatsumisago90,Bohmer92}. { Interesting works using density functional theory~\cite{Hall03,Micoulaut03a} investigated the relationship between structure and fragility, but they do not capture this non-monotonicity}. 

Recent observations \cite{Trachenko00,OHern03,Chen08,Chen10,Ghosh10} and theory \cite{Wyart05a,Wyart05b,Xu07,Brito09,Souza09,Souza09a,DeGiuli14, DeGiuli14b,Degiuli15,Franz15b} indicate that  in various amorphous materials, the presence of soft elastic modes is regulated by the proximity of the rigidity transition, linking  evidence (I) and (II). To rationalize this connection,  we have introduced a frozen elastic network model that bridges the gap between network elasticity and geometry on one hand, elasticity and the thermodynamics and dynamics of liquids on the other~\cite{Yan13}. This model incorporated the following aspects of supercooled liquids: ({\it i}) particles interact with each other with interactions that can greatly differ in strength, such as the covalent bonds and the much weaker Van der Waals interaction found in network glasses. ({\it ii}) Neighboring particles can organize into a few distinct local configurations.  ({\it iii}) The choices of local configurations are coupled at different location in space via elasticity. These features were modeled using a random elastic network whose topology was frozen, as illustrated in Fig.~\ref{model}. The possibility for local configurations to change was incorporated by letting each spring switch between two possible rest lengths. 
Despite its simplicity, this model recovered (I) and (II). In particular, it reproduced the non-monotonic variance of the jump of specific heat and the fragility with the coordination  $z$ of the network: they are extremal at $z_c=2d$ ($d$ is the spatial dimension),  where a rigidity transition occurs. This model could be solved analytically, and it led to the view that near the rigidity transition, the jump of specific heat is small because frustration vanishes: most directions in phase space do not cost energy, and thus do not contribute to the specific heat.

\begin{figure}[h!]
\includegraphics[width=1.0\columnwidth]{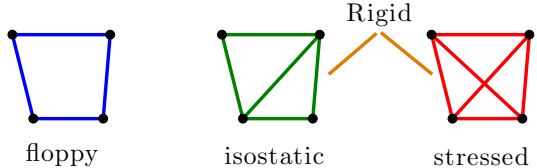}
\caption{\small{(Color online) Illustration of rigidity transition. Blue, green, and red color the floppy, isostatic, and stressed clusters, respectively.}}\label{isostatic}
\end{figure}

This is a novel explanation for a long-standing problem, and it is important to confirm that this view is robust
when more realism is brought into the model. In particular, the model used frozen disorder to describe elasticity,
whereas it is itself a dynamical property in liquids, where there cannot be any frozen disorder. 
The thermal evolution of the topology of the contact network and its effects on rigidity transition were also not addressed. 
 A network is rigid when an imposed global strain induces stress, and the rigidity can be achieved topologically by adding  constraints~\cite{Maxwell64}, see Fig.~\ref{isostatic} for an illustration in a small network. The network is said to be self-stressed if some of the constraints are redundant, removing those leaves  the network rigid. 
Three scenarios of rigidity transition have been extensively studied in the literature~\cite{Yan14,Ellenbroek15} (but see Ref. \cite{Moukarzel13} for a recent fourth proposition). 
Spatial fluctuations of coordination  are important in the first two. The {\it rigidity percolation} model \cite{Jacobs98,Duxbury99,Feng84,Jacobs95} assumes that bonds are randomly deposited on a lattice. Fluctuations  lead to over-constrained (self-stressed) clusters  even when the average coordination number is not sufficient to make the whole network rigid.
This  model corresponds to the infinite temperature limit.  To include these effects, self-organized network models were introduced \cite{Thorpe00,Chubynsky06,Briere07,Micoulaut03}, where overconstrained regions are penalized. A surprising outcome of these models is the emergence of a rigidity window:  rigidity emerges at a small coordination number before the self-stress appears 
(even in the thermodynamic limit). 
Finally, in the {\it mean-field  or jamming scenario}, fluctuations of coordinations are limited.  Similar to the simple picture in Fig.~\ref{isostatic}, the rigidity, and the stress appear at the same $z_c$ in the thermodynamic limit. The rigid cluster at $z_c$ is not fractal and is similar to that of packings of repulsive particles. The model of Ref. \cite{Yan13} assumed that networks were of this last type.

Recently, we have introduced adaptive elastic network models \cite{Yan14}, where the topology of the network is free to evolve to lower its elastic energy as the system is cooled. We found that as soon as weak interactions are present, the network of strong interactions becomes mean-field like at low temperature.  However, the thermodynamic properties  were not studied to test the robustness of the thermodynamic predictions of Ref. \cite{Yan13} relating structure to the jump of specific heat. In this work, we directly show numerically and theoretically that the prediction for the jump of specific heat is essentially identical in adaptive and frozen elastic network models. Section II describes the adaptive network models. Section III presents the numerical results of the model, while Section IV gives the explicit derivation of the thermodynamic properties, developing an approximation scheme to deal with the temperature-dependence of the number of over-constraints in the system, treating them as an ideal gas.

\section{II. Model}

\begin{figure}[h!]
\includegraphics[width=1.0\columnwidth]{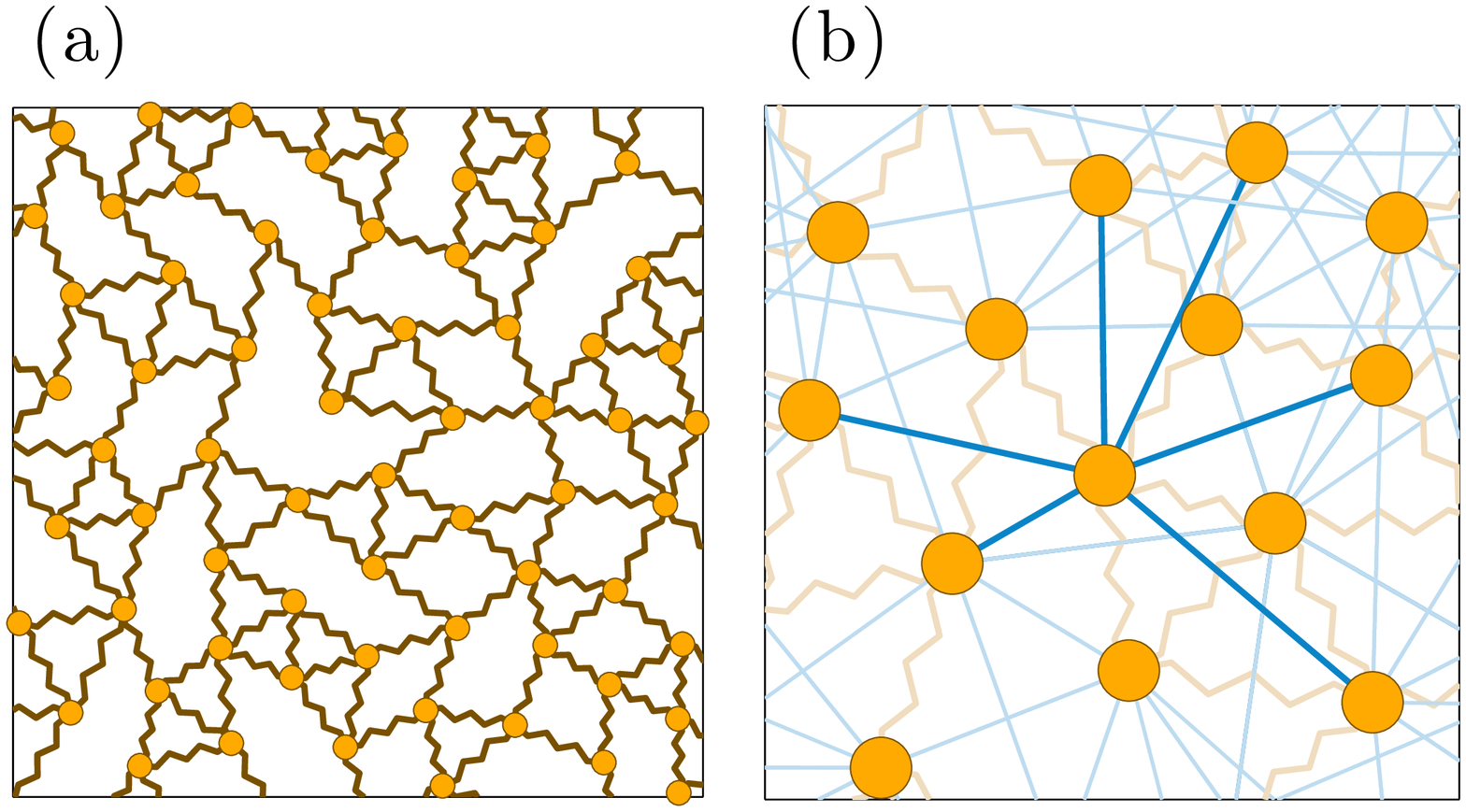}
\includegraphics[width=1.0\columnwidth]{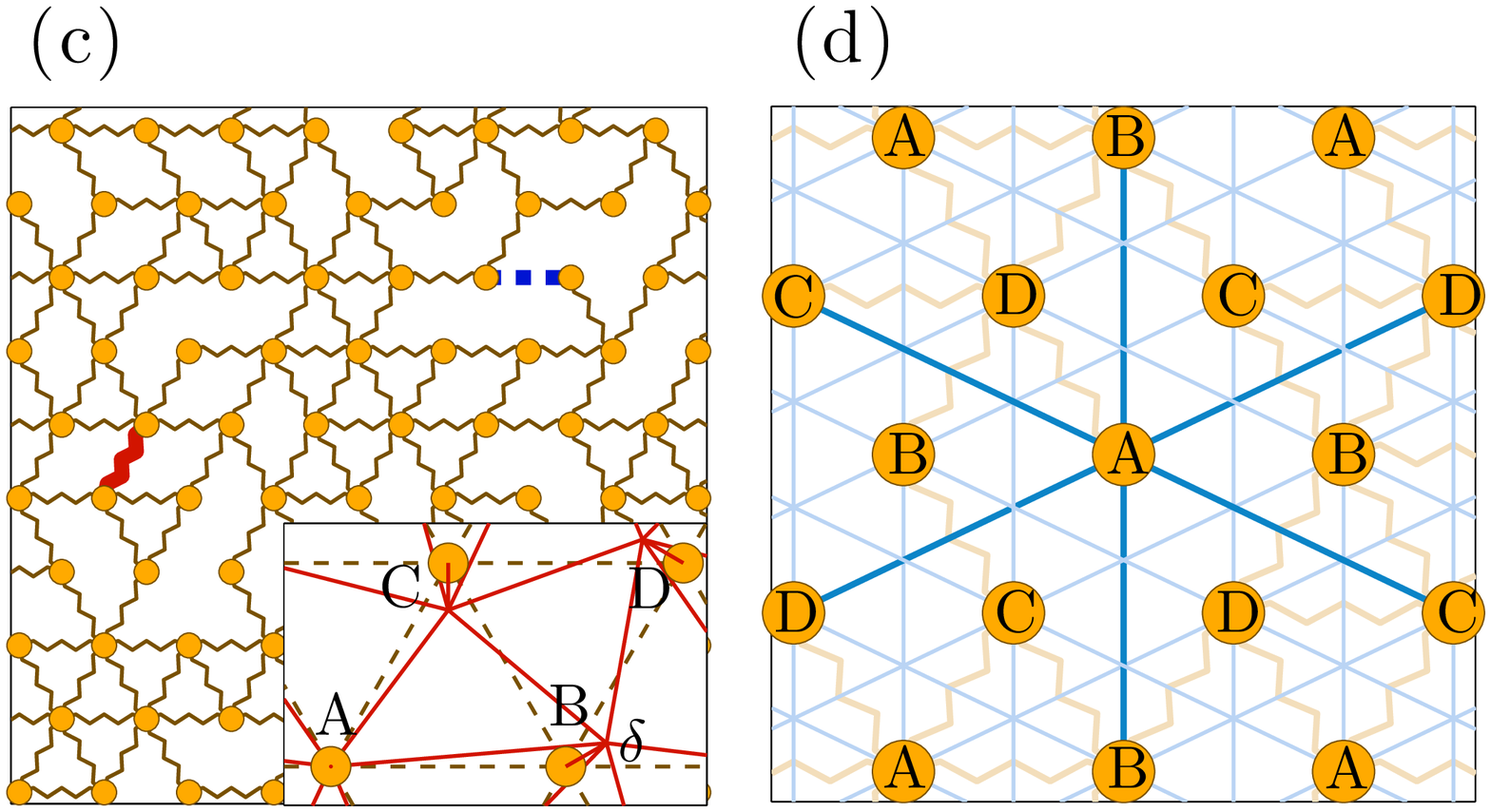}
\caption{\small{(Color online) (a) and (b) Illustration of the frozen network model~\cite{Yan13}; (c) and (d) illustrate the adaptive network model~\cite{Yan14}. In the latter case, the triangular lattice is systematically distorted in a unit cell of four nodes shown in the inset of (c). We group nodes by four, labeled as, A, B, C, and D in Fig.~\ref{model}. One group forms the unit cell of the crystalline lattice. Each cell is distorted identically in the following way: node A stays,  while nodes B, C, and D move by a distance $\delta$, B along the direction perpendicular to BC, C along the direction perpendicular to CD, and D along the direction perpendicular to DB. $\delta$ is set to $0.2$ with the lattice constant as unity. Weak springs connecting (b) six nearest neighbors without strong springs and (d) six next-nearest-neighbors are indicated in straight cyan lines, emphasized for the central node.  (c) Illustration of an allowed step, where the strong spring in red relocates to a vacant edge indicated  by a dashed  blue line.}}\label{model}
\end{figure}

In our  model degrees of freedom are springs, which are poly-disperse and can move  on a lattice. The lattice is built using a triangular lattice with periodic boundary conditions, see Fig.~\ref{model}(c), with a slight regular distortion to { minimize the non-generic presence of zero modes that occurs when straight lines are present}, as illustrated in the inset of Fig.~\ref{model}(c).    
Polydisperse and mobile ``strong" springs of identical stiffness $k$ connect the nearest neighbors on the lattice and model the covalent constraints. 
We model weak Van der Waals interactions with ``weak" and stationary  springs of stiffness $k_{\rw}\ll k$ adding to all next-nearest-neighbors on the triangular lattice, illustrated in Fig.~\ref{model}(b). We introduce a control parameter $\alpha\equiv(z_{\rw}/d)(k_{\rw}/k)$ to characterize the relative strength of the weak interactions, where the spatial dimension is $d=2$ and the  number of weak constraints per node is chosen $z_{\rw}=6$.

 The number of ``covalent'' springs $N_s$, equivalent to the coordination number $z\equiv2N_s/N$ ($N$ is the number of nodes in the lattice), is also a dimensionless control parameter. 
For a given { $\delta z\equiv z-z_c$}, the valid configurations are defined by the locations of the $N_s$ springs, indicated as $\Gamma\equiv\{\gamma\leftrightarrow\langle i,j\rangle\}$, where the Greek index $\gamma$ labels springs and the Roman indices $\langle i,j\rangle$ label the edges on triangular lattice between nodes $i$ and $j$. We introduce the { occupation of an edge:} $\sigma_{\langle i,j\rangle}=0$ if there is no strong spring on the edge $ij$, and $\sigma_{\langle i,j\rangle}=1$ if there is one. 
If $r_{\langle i,j\rangle}$ denotes the geometric length between nodes $i$ and $j$ on the lattice, we assume that the spring $\gamma$ has a rest length $l_{\gamma}=r_{\langle i,j\rangle}+\epsilon_{\gamma}$, where the mismatch $\epsilon_{\gamma}$ is a feature of a given spring. $\epsilon_{\gamma}$ are sampled independently from a Gaussian distribution with mean zero and variance $\epsilon^2$, which thus characterizes the polydispersity of the model. $k\epsilon^2$ is set to unity as the natural energy scale.

\begin{figure}[h!]
\includegraphics[width=1.0\columnwidth]{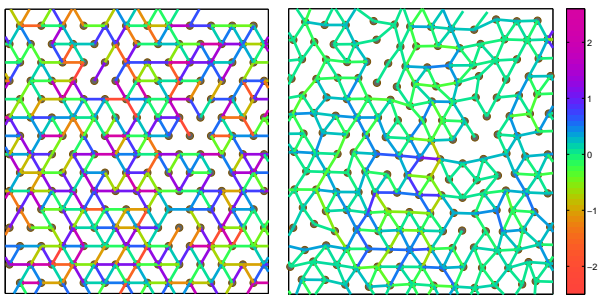}
\caption{\small{(Color online) Illustration of configuration energy of the adaptive network model ($\delta z=0.27$). { Solid lines are springs, colored according to their extensions: from red to purple, the springs go from being  stretched to being compressed,  with spring extensions shown in the unit of $\epsilon$. Left: Nodes sit at lattice sites, so the color shows the rest length mismatches of the springs $\{\epsilon_{\gamma}\}$. Right: Nodes are relaxed to mechanical equilibrium. Most links appear in green, indicating that most of the elastic energy is released. The configuration energy is defined by the residual energy.}}}\label{relax}
\end{figure}

The energy of an inherent structure is denoted $\mh(\Gamma)$. The configuration $\Gamma$ is sampled with probability proportional to $\exp(-\mh(\Gamma)/T)$ in the liquid phase, with $k_B=1$. Temperature $T$ serves as a third dimensionless control parameter. $\mh(\Gamma)$ is defined as the remaining energy once the nodes of the network are allowed to relax to mechanical equilibrium:
\begin{multline}
\label{e1}
\mh(\Gamma)=
\min_{\{ {\vec R}_i\}} \left\{\sum_\gamma \frac{k}{2} \left[||{\vec R}_i-{\vec R}_j||-l_\gamma\right]^2\right.\\\left.+\sum_{\langle i,j\rangle_2}\frac{k_{\rm w}}{2}\left[||{\vec R}_i-{\vec R}_j||-r_{\langle i, j\rangle_2}\right]^2\right\}
\end{multline}
where $\vec{R}_i$ is the position of particle $i$ and $\langle i, j\rangle_2$ labels the next-nearest neighbors. The minimal energy can be calculated by steepest decent as illustrated in Fig.~\ref{relax}, but this is computationally expensive. Instead, we approximate the elastic energy in the linear response range, setting that $\epsilon^2\ll1$~\footnote{We have tested the validity of the linear approximation: the energy difference from the steepest decent results keeps below 3\% for $\epsilon<0.02$.}. The above minimization expression Eq.(\ref{e1}) could then be written as,
\be
\mh(\Gamma)=\frac{k}{2}\sum_{\Gamma}\epsilon_{\langle i,j\rangle}\mg_{\langle i,j\rangle,\langle l,m\rangle}\epsilon_{\langle l,m\rangle}+o(\epsilon^3)
\label{hamiltonian}
\ee
where $\epsilon_{\langle i,j\rangle}=\epsilon_{\gamma}$ when spring $\gamma$ connects $i$ and $j$. 
The coupling matrix $\mg=\mpp-\ms(\ms^t\ms+\frac{k_{\rw}}{k}\ms_{\rw}^t\ms_{\rw})^{-1}\ms^t$, derived in our previous works~\cite{Yan13,Yan14} (or see Appendix Sec.~A), is a product of the structure matrix $\ms$ and its transpose $\ms^t$, the structure matrix of the weak spring network $\ms_{\rw}$, and $\mpp$ the projection operator of the triangular lattice onto occupied edges. The structure matrices $\ms$ and $\ms_{\rw}$ describe the topology of the networks of strong and weak springs: if neighbor nodes $i$ and $j$ are connected, the change of the distance between $i$ and $j$, $\delta r_{\langle i,j\rangle}=\ms_{\langle i,j\rangle,i}\cdot\delta\vec{R}_i+\ms_{\langle i,j\rangle,j}\cdot\delta\vec{R}_j+o(\delta\vec{R}^2)$, due to displacements of nodes $\delta\vec{R}$. We point out that as the weak network is fixed, $\ms$ and thus $\mg$ depend only on the network topology of strong springs, but not on the mismatches $\epsilon_{\gamma}$. 

Our model is a generalization of on-lattice network models: setting the interaction strength control parameter $\alpha=0$, it naturally recovers the randomly diluted lattice model~\cite{Jacobs95} when $T=\infty$. It is also related to the self-organized lattice model~\cite{Thorpe00,Chubynsky06}, which postulates that elastic energy is linearly proportional to the number of redundant constraints~\cite{Thorpe00,Barre09}. We will find that this assumption holds true for $\alpha=0$ and $T\ll1$.
However, the existence of weak interactions among sites means that in real physical systems $\alpha>0$. This turns out to completely change the physics, an effect that our model can incorporate.

\section{III. Numerics}
We implement a Monte Carlo simulation to sample the configuration space of the model, with $10^6$ Monte Carlo steps at each $T$. At each step, a potential configuration is generated by a Glauber dynamics - moving one randomly chosen spring to a vacant edge, as illustrated in Fig.~\ref{model}(c). 
We numerically compute the elastic energy of the proposed configuration using Eq.(\ref{hamiltonian}): calculating the structure matrix $\ms$ and then the corresponding $\mg$. On computing $\mg$, the matrix { inversion, $(\ms^t\ms+\frac{k_{\rw}}{k}\ms_{\rw}^t\ms_{\rw})^{-1}$,} is singular when the network contains floppy structures, which do not appear except when $k_{\rw}=0$. When $\alpha=0$, we implement the ``pebble game'' algorithm~\cite{Jacobs97} to identify the over-constrained sub-networks, and then do matrix division in the subspace, as the isostatic and floppy regions store no elastic energy after relaxation. 
We have found little finite size effect by varying the system size from $N=64$ to $N=1024$ nodes in the triangular lattice. In the following, we present our numerical results of networks with $N=256$ nodes, averaged over 50 realizations of random mismatches if not specified.

\subsection{A. Dynamics}
\begin{figure}[h!]
\centering
\includegraphics[width=.9\columnwidth]{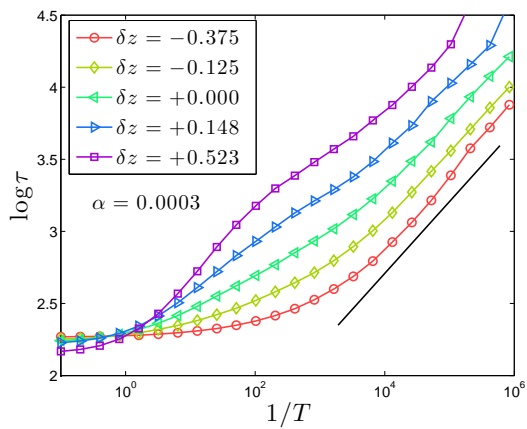}
\caption{\small{(Color online) Relaxation time $\tau$ in log-scale versus inverse temperature $1/T$ for different coordination numbers $\delta z$ and $\alpha=0.0003$. The solid black line indicates a power law relation between $\tau$ and $T$: $\tau\sim T^{-1/2}$.}}\label{logt}
\end{figure}

We investigate the dynamics by computing the correlation function $C(t)=\frac{1}{N_{s}(1-N_s/3N)}(\langle\sigma(t)|\sigma(0)\rangle-N_s^2/3N)$, { where $|\sigma(t)\rangle$ is the vector indicating the occupation  of all edges at time $t$. The correlation $C(t)$} decays from one to zero at long time scales.  We define the relaxation time $\tau$ as the time $C(\tau)=1/2$, and the numerical results of $\tau$ as a function of temperature $T$ for several different coordination numbers are shown in the Fig.~\ref{logt}. 

We find that the implemented dynamics is not glassy. The relaxation time increases as a power law of the temperature $T^{-0.5}$, even much slower than a strong glass that would display an Arrhenius behavior $\log_{10}\tau\propto1/T$.  This result is very surprising because the frozen elastic network model we studied earlier was glassy (its fragility was similar to that of network liquids). Despite being dynamically very different, these two models are almost identical as far as thermodynamics is concerned, as we will see below. It could be that the lack of glassiness comes from our choice of Monte-Carlo where springs can try other locations anywhere in the system~\cite{Grigera01a}. 

%

\begin{figure}[t!]
\includegraphics[width=1.0\columnwidth]{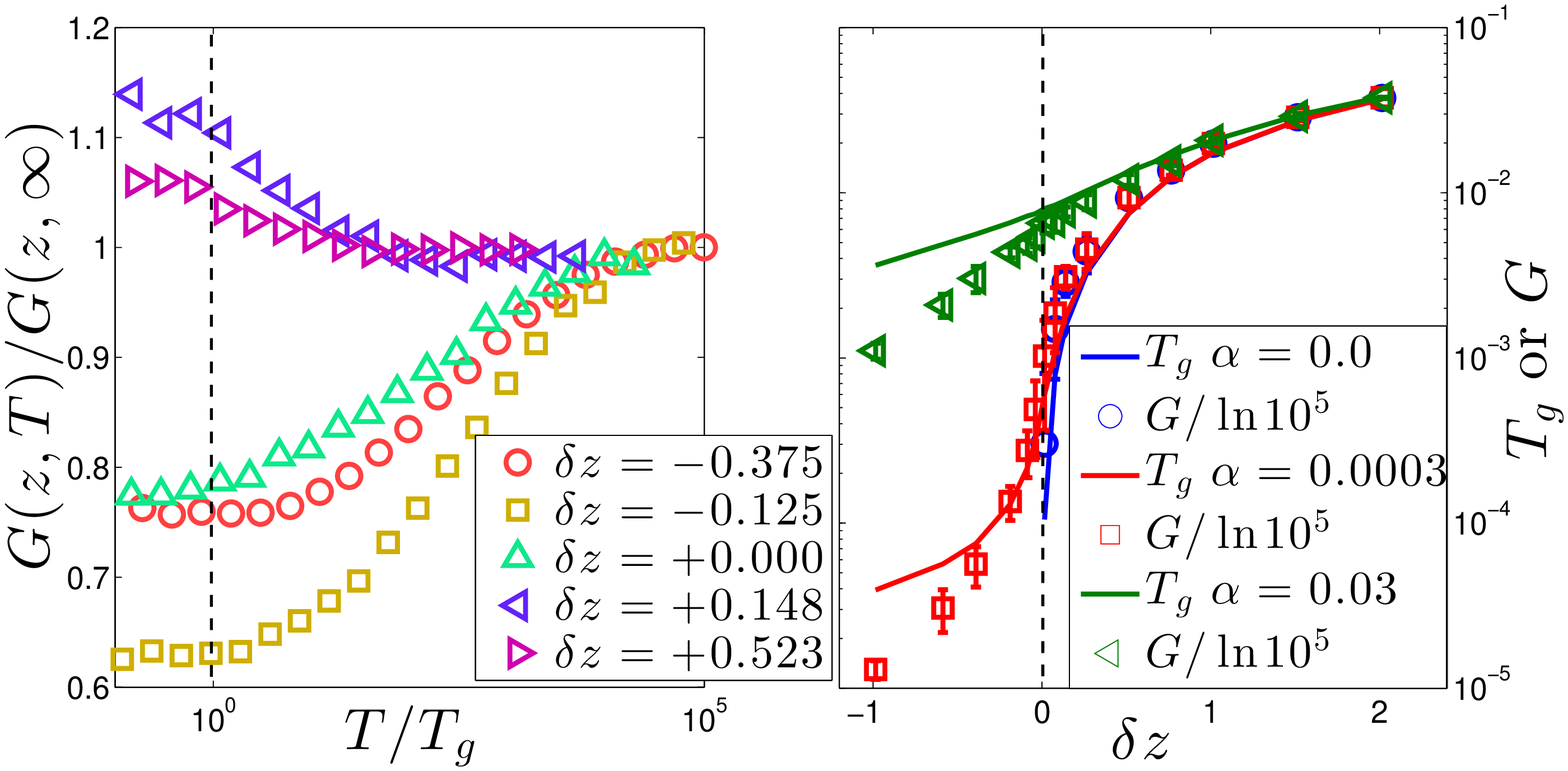}
\caption{\small{(Color online) Left: Shear modulus of adaptive networks at temperature $T$ rescaled by $G$ at $T=\infty$ $G(z,T)/G(z,\infty)$, $\alpha=0.0003$. The temperature $T$ is rescaled by $T_g$. Right: Correlation between transition temperature $T_g$ and shear modulus $G$ in the frozen network model~\cite{Yan13}. 
}}\label{GT}
\end{figure}

To compare the thermodynamics of these models we now need to define an effective glass temperature $T_g$ (even if we do not see a real glass transition). { We do that by using the  empirical Lindemann criterion~\cite{Lindemann10} according to which an amorphous solid melts when the standard deviation $\langle \delta R^2\rangle^{1/2}$ of particles' displacements is greater than a fraction $c_L$ of the particle size $a$. The coefficient $c_L$ must depends on the quench rate $q$, since this is also the case for $T_g$. This dependence is logarithmic, because the dependence of relaxation time on temperature in experimental glass formers is  at least exponential (for typical experimental quench rate in supercooled liquids, $c_L\approx 0.15$   ~\cite{Nelson02})}.  We can estimate this standard deviation via the elastic modulus if we treat the glass as a continuum $\langle \delta R^2\rangle\sim T/G a$ where $G$ is the instantaneous shear modulus of the structure~\cite{Dyre06}, we thus get $T_g\propto Ga^3/\ln(1/q)$. We set the lattice length $a$ in our model to unity. 

We measure the shear modulus averaging over configurations at given temperatures, shown in the left panel of Fig.~\ref{GT}. 
Practically, we choose { $T_g=\langle G\rangle_{T_g}/\ln(1/10^3 q)$, where the cooling rate $q$ is defined as the inverse of  the number of Monte Carlo steps performed at each temperature in the model}. $\langle\bullet\rangle_{T_g}$ is the mean value at temperature $T_g$. The prefactor in this definition of $T_g$ does not affect qualitatively { our conclusions, but for this pre-factor the definition of $T_g$ in the frozen model \cite{Yan13} is essentially identical to the dynamical definition used in \cite{Yan13}, as shown in the right panel of Fig.~\ref{GT} by lining up $G$ and $T_g$.  The specific values of $T_g$ following that definition are shown in the inset of the bottom panel of Fig.~\ref{cp}, they correspond to $T_g=\langle G\rangle_{T_g}/\ln(10^3 )$ in the present model, and $T_g=\langle G\rangle_{T_g}/\ln(10^5 )$ in the frozen network model~\cite{Yan13}, which is simpler to simulate and can thus be equilibrated longer.} 

\begin{figure*}[ht]
\includegraphics[width=2.0\columnwidth]{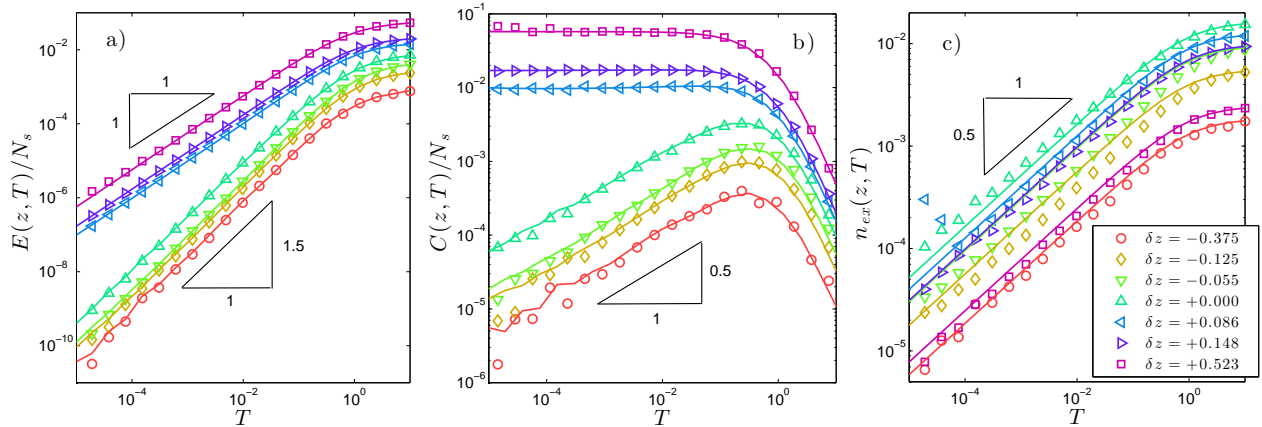}
\caption{\small{(Color online) Thermodynamics of the adaptive network model without weak constraints $\alpha=0$. (a) Energy $E/N_s$ {\it vs} temperature; (b) Specific heat $C/N_s$ {\it vs} temperature; (c) Excess number density of redundant constraints $n_{\rm ex}$ extracted using the pebble game algorithm {\it vs} temperature. Symbols are numerical data, solid lines are theoretic predictions.
}}\label{thermo0}
\end{figure*}

\subsection{B. Specific heat}

\begin{figure}[h!]
\centering
\includegraphics[width=.9\columnwidth]{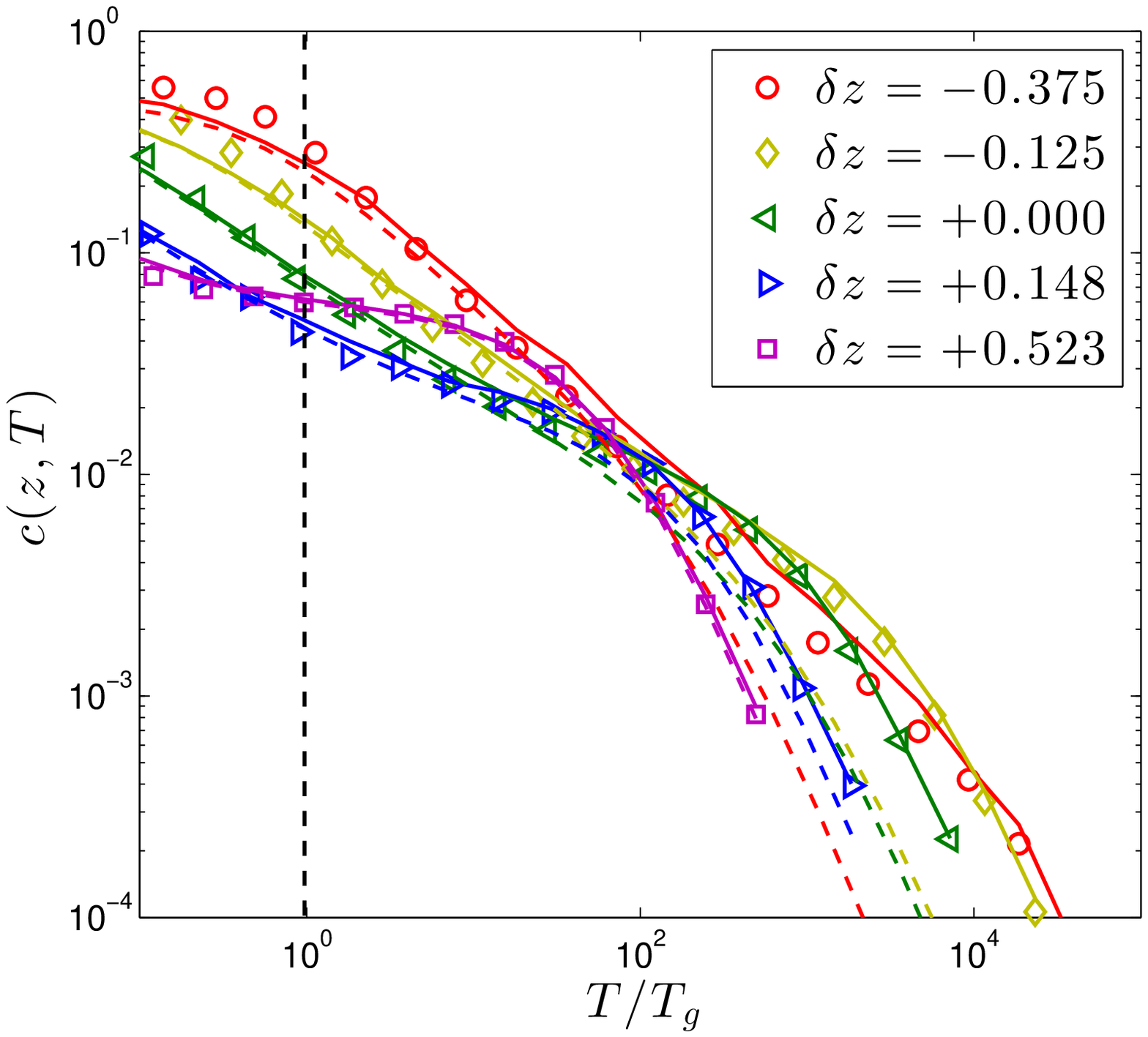}\\
\includegraphics[width=.9\columnwidth]{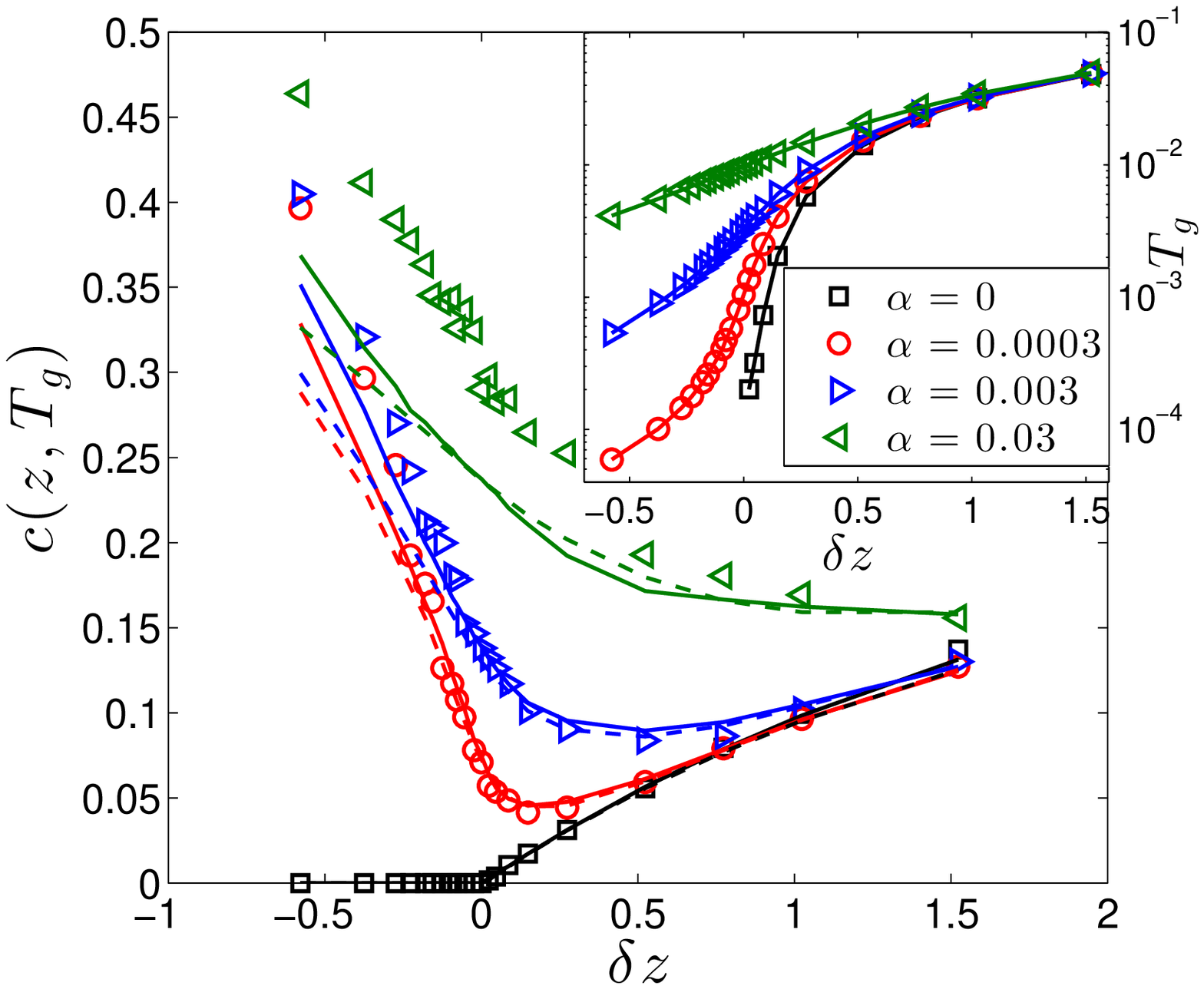}
\caption{\small{(Color online) Top: Specific heat $c(z,T)$ {\it vs} scaled temperature $T/T_g$ for networks with average coordination numbers near and away from the isostatic on both floppy and rigid sides. The strength of the weak constraints is given by $\alpha=0.0003$. Bottom: Specific heat at temperature $T_g$, $c(z,T_g)$, {\it vs} coordination number $\delta z$ for $\alpha=0,\ 0.0003,\ 0.003,\ 0.03.$  The inset shows the transition temperature $T_g$ for different $z$ and $\alpha$. Symbols are numerical results, and lines are theoretical predictions: dashed lines are for frozen network model and solid lines are for the new model derived in section IV.}}\label{cp}
\end{figure}

The specific heat data shown in Figs.~\ref{thermo0} and \ref{cp} are our central numerical results. 
The energy $E=\langle \mh\rangle$ is  obtained using a time-average over Monte Carlo steps, and is shown in Fig.~\ref{thermo0}(a). The specific heat is calculated as its derivative $c\equiv\frac{1}{N_s}\rd E/\rd T$, and is shown   versus $T$ for several coordination numbers when $\alpha=0$ in Fig.~\ref{thermo0}(b)  and $\alpha=0.0003$ in the top panel of Fig.~\ref{cp}. When $\alpha=0$, the specific heat increases as temperature decreases for networks with $\delta z>0$ while it meets a maximum at $T_a\sim1$ and decreases under cooling when $T<T_a$ if $\delta z\leq0$. By contrast, the specific heat increases under cooling close to the transition temperature for all coordination numbers when $\alpha>0$. In addition, when $T\lesssim\alpha$, $c\to0.5$.
All these results are qualitatively identical to our previous frozen model. 

To define the jump of the specific heat at the glass transition, we simply measure the specific heat at our glass transition $T_g$ defined above.  This definition is natural, since in a real glassy system, below $T_{g}$ the liquid is  essentially frozen in an inherent structure, and the contribution to the specific heat from configurational entropy (i.e. the bottom energy of inherent structures) vanishes.  

Our central numerical result is shown in the bottom panel of Fig.~\ref{cp}:  $c(T_g)$ varies non-monotonically with the coordination number $z$ when $\alpha>0$. When the network of strong springs is poorly coordinated $\delta z\lesssim0$, $c(T_g)$ decreases as $z$ increases; When the strong network gets better coordinated $\delta z\gtrsim0$, $c$ gradually changes to increase with $z$; $c$ is minimal at the proximity of the rigidity transition $z_c$ for finite $\alpha$. These numerical results are very similar to empirical observations, see Point (II) in the introduction.
Our data are in fact very similar to that of the frozen model, which essentially follows the dotted lines in Fig.~\ref{cp}. 

\subsection{C. Number of redundant constraints $R$}
When $\alpha=0$ and $T\to0$, the specific heat is simply proportional to $R$, as shown in Fig.~\ref{thermo0}(b). This number is fixed, $R=N\delta z/2$, in the frozen network models. It varies in the adaptive network model and depends on the temperature. As the Maxwell counting gives the minimal number of redundant constraints of a network, we can define an excess number of redundant constraints
\be
\label{nex}
n_{\rm ex}\equiv \frac{1}{N_s}\left(R-\frac{N\delta z}{2}\Theta(\delta z)\right),
\ee
where $\Theta(x)$ is the Heaviside step function. $n_{\rm ex}$ counts the average number of redundant constraints, additional to the Maxwell counting. 
This excess number of redundant constraints decreases monotonically to zero under cooling. When $\alpha=0$, $n_{\rm ex}$ is proportional to $\sqrt{T}$ in the adaptive network model at low temperature, shown in Fig.~\ref{thermo0}(c).

\begin{figure}[h!]
\centering
\includegraphics[width=.9\columnwidth]{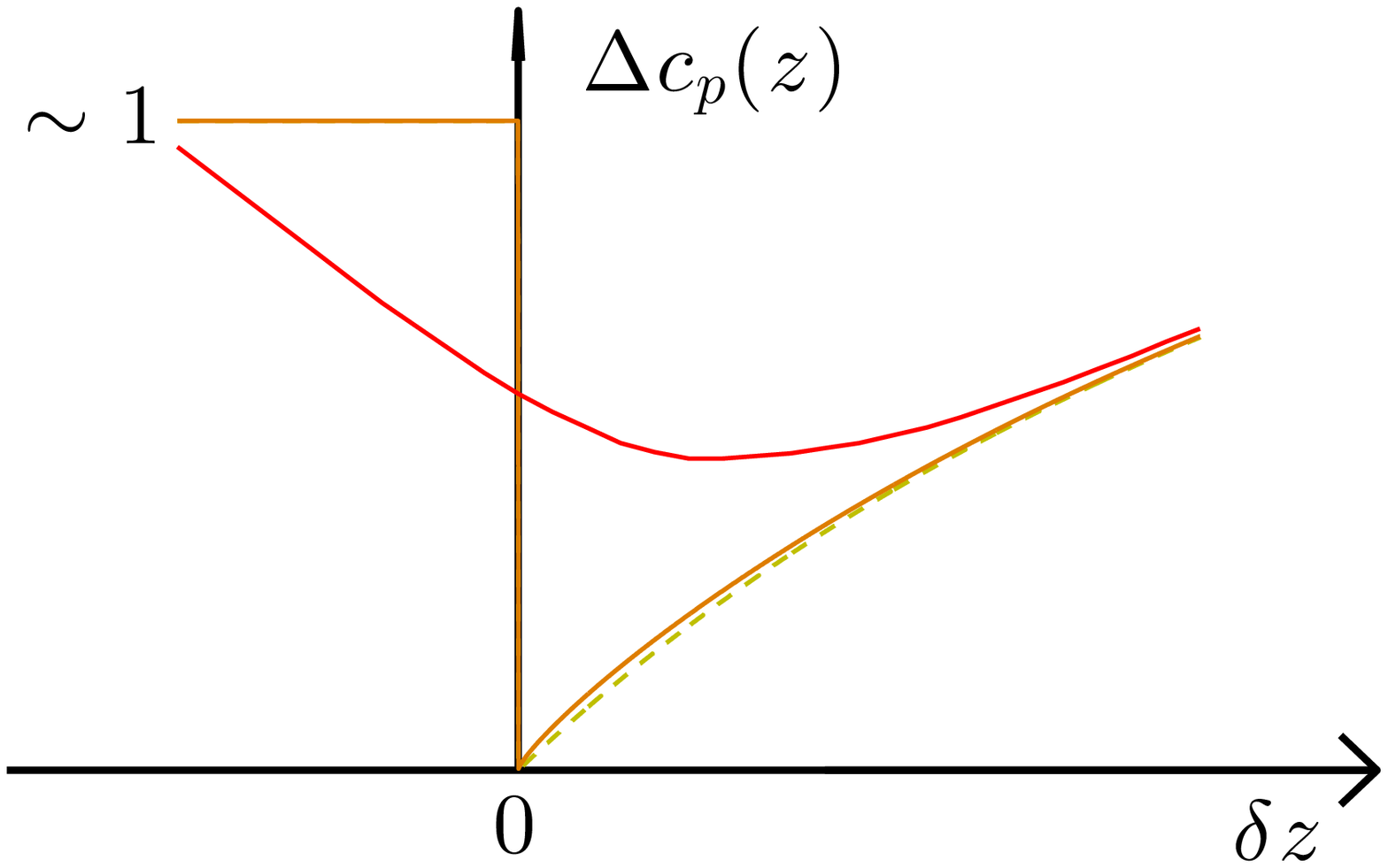}
\caption{\small{(Color online) Theoretical predictions for the jump of specific heat. For vanishingly weak springs $\alpha\rightarrow 0$, it is predicted that the jump is essentially constant for $z<z_c$ and then drops to zero a $z_c$. For larger $z$, it behaves as $z-z_c$. As $\alpha$ grows this sharp curve becomes smooth, but a  minimum is still present near $z=z_c$.  }}\label{theory}
\end{figure}

\section{IV. Theory}

As illustrated in Fig.~\ref{theory}, in the frozen elastic model we found that as $\alpha\rightarrow 0$, $c$ converges to a constant if $z<z_c$,
whereas it behaves as $z-z_c$ for $z>z_c$. As $\alpha$ is increased, the discontinuous behavior becomes smooth and looks similar to experimental data. We seek to derive these same features in the adaptive network models.

%

\subsection{A. Thermodynamics}
For simplicity, we consider the annealed free energy $\mf_{\rm ann}=-T\ln\overline{\mz}$. It is exact in the random energy model~\cite{Derrida81} above the ideal glass transition~\cite{Mezard09} and we find it to be a good approximation of $\overline{\mf}$ in our models~\cite{Yan13}. 
The over-line implies an average over disorder $\epsilon$, 
\be
\overline{\mz}=\overline{\sum_{\{\sigma\}}\sum_{\text{perm}[\gamma]}\exp[-\mh(\Gamma)/T]}
\label{partfunc}
\ee
where a given configuration $\Gamma$ is characterized by $\{\sigma\}$  indicating which  edges are occupied on  the triangular lattice, 
and $\text{perm}[\gamma]$ labels the possible permutations of springs' rest lengths. 

We first average over the quenched randomnesses. Using the linear approximation Eq.(\ref{hamiltonian}) and the Gaussian distribution $\rho(\epsilon_{\gamma})=\frac{1}{\sqrt{2\pi\epsilon^2}}e^{-\epsilon_{\gamma}^2/2\epsilon^2}$,
\be
\overline{\mz}=\sum_{\{\sigma\}}\left(\frac{Nz}{2}\right)!\exp\left[-\frac{1}{2}\tr\ln\left(\mi+\frac{\mg(\{\sigma\})}{T}\right)\right]
\label{partz}
\ee
The factorial comes from $N_s!=\sum_{\text{perm}[\gamma]}{\bf 1}$ as $\mg$ is independent of the permutation. $\mi$ is a $3N\times 3N$ identity matrix; each component corresponds to an edge on the lattice. 
To compute the trace in the exponent, we first make the approximation that the  weak springs are weak and numerous $\ms_{\rw}^t\ms_{\rw}\approx\frac{z_{\rw}}{d}\mi_{Nd\times Nd}$, which corresponds to the highly connected limit $z_{\rw}\to\infty$ and finite $\alpha$. We can then decompose the coupling matrix $\mg\approx\mpp-\ms(\ms^t\ms+\alpha\mi)^{-1}\ms^t$ as ~\cite{Yan13}:
\be
\mg(\{\sigma\})=\sum_{p(\{\sigma\})}|\psi_{p}\rangle\langle\psi_p|+\sum_{\omega(\{\sigma\})>0}\frac{\alpha}{\omega^2+\alpha}|\psi_{\omega}\rangle\langle\psi_{\omega}|
\label{coupmat}
\ee
where $p$ labels the vectors $|\psi_p\rangle$ satisfying $\ms^t|\psi_p\rangle=0$ (i.e. a basis for the kernel of $\ms^t$), and where the $|\psi_{\omega}\rangle$ satisfy $\ms\ms^t|\psi_{\omega}\rangle=\omega^2|\psi_{\omega}\rangle$. The number of redundant directions is $\sum_{p}{\bf 1}=N_s-(Nd-F)\equiv R$. Note that  $\tr\mpp=N_s$, $Nd-F$ gives the number of frequencies $\omega$, and $F$ counts the number of floppy modes. The modes  $|\psi_p\rangle$, $|\psi_{\omega}\rangle$, $R$, and $\omega$ depend on occupation $\{\sigma\}$. As the $|\psi\rangle$'s are orthonormal, the trace in Eq.(\ref{partz}) gives 
\begin{widetext}
\be
\overline{\mz}=\left(\frac{Nz}{2}\right)!\sum_{n_r,D(\omega)}\exp\left[N_s\left(s(n_r,D(\omega))-\frac{n_r}{2}\ln(1+\frac{1}{T})-\frac{1-n_r}{2}\int\rd\omega D(\omega)\ln(1+\frac{1}{T}\frac{\alpha}{\omega^2+\alpha})\right)\right],
\label{part}
\ee
\end{widetext}
where $s(n_r,D(\omega))\equiv\frac{1}{N_s}\ln\sum_{\{\sigma\}}{\bf 1}_{R,D(\omega)}$ is configurational entropy density with given number of redundant constraints $n_r\equiv R/N_s$ and density of vibrational modes, $D(\omega)$, satisfies $(1-n_r)\int\rd\omega D(\omega)\equiv\lim_{N\to\infty}\frac{1}{N_s}\sum_{\omega>0}$.

\subsection{B. No weak interactions}

\begin{figure}[h!]
\includegraphics[width=1.0\columnwidth]{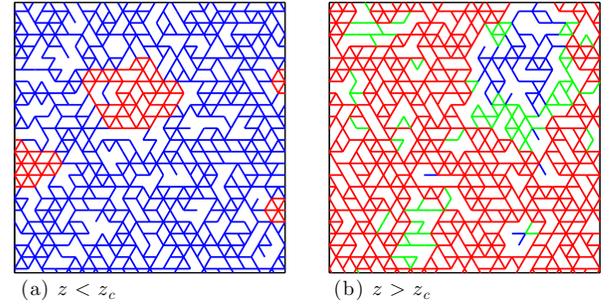}
\caption{\small{(Color online) (a) $z<z_c$, localized redundant constraints (red) in a floppy sea (blue); (b) $z>z_c$ localized floppy modes (blue) in a rigid sea (red and green).}}\label{defect}
\end{figure}

Neglecting the weak constraints $\alpha=0$, the last term in the exponential vanishes and the summation over states with given density of states can be absorbed into the entropy, which then depends only on the number of redundant constraints. 
\be
\overline{\mz}=\left(\frac{Nz}{2}\right)!\sum_{n_r}e^{N_s[s(n_r)-\frac{n_r}{2}\ln(1+\frac{1}{T})]}
\label{part0}
\ee

We propose an ideal-gas picture of ``defects''  to find an approximation form of the entropy $s(n_r)$. 
When the coordination number is very small $z<z_c$ and the network is mostly floppy, redundant constraints are defects localized in rigid islands. Similarly, when the coordination number is very large $z>z_c$ with most regions of the network rigid, there are localized floppy modes in regions where there are negative fluctuations of coordination number, which we again described as defects, see illustration in   Fig.~\ref{defect}. The number of such floppy modes is equal to the number of additional over-constrained in the rigid cluster. The entropy gains from having these defects. Assuming that such defects are independent, we approximate the entropy by that of an ideal  gas: 
\be
s(n_{\rm ex})\approx s(0)-n_{\rm ex}\ln\frac{n_{\rm ex}}{en_0(z)}
\label{s}
\ee
where $n_{\rm ex}$ is the excess number of redundant constraints defined in Eq.(\ref{nex}) and is thus counting the number of defects. $s(0)$ is the entropy density of the states with a minimal number of redundant constraints (i.e. they satisfy the Maxwell counting); and $n_0(z)$ is the excess number of redundant constraints at $T=\infty$. Both $s(0)$ and $n_0$ depend only on $z$ and the lattice structure. This form of Eq.(\ref{s}) fails when the assumption of independent ``defects'' breaks down, as must occur near the rigidity transition. However,  our numerical results indicate that this approximation is very accurate, we see deviations only for $|\delta z|\lesssim0.1$. 

\begin{figure}[h!]
\includegraphics[width=1.0\columnwidth]{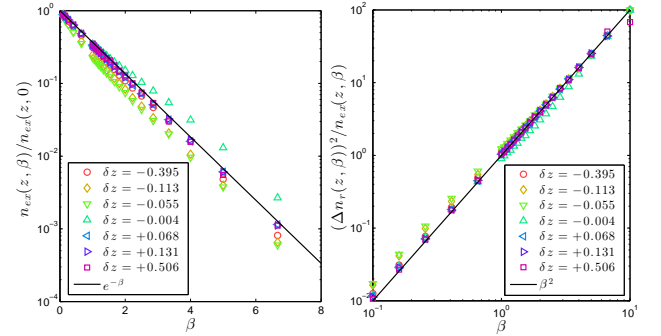}\\
\caption{\small{(Color online) Left: Excess number density of redundant constraints  $n_{\rm ex}(z,\beta)$.  Right: Fluctuation of the number density of redundant constraints  $(\Delta n_r)^2$. The solid black lines show the predictions from the approximate entropy Eq.(\ref{s}). 
}}\label{snr}
\end{figure}

We numerically test the formula Eq.(\ref{s}) for a triangular lattice. 
The configurations with $R$ redundant constraints are weighted by $e^{-\beta R}$ for different values of the parameter $\beta$. From Eq.(\ref{s}), the mean and variance of the excess number density of redundant constraints, $n_{\rm ex}$, satisfy the following formulas: 
\begin{subequations}
\begin{equation}
\beta\equiv\frac{\partial s}{\partial n_{\rm ex}}\ \Rightarrow n_{\rm ex}(z,\beta)=n_0(z)e^{-\beta}
\label{nbeta}
\end{equation}
\begin{equation}
\Delta n_{\rm ex}^2(z,\beta)=-\beta^2\frac{\partial}{\partial\beta} n_{\rm ex}(z,\beta)=\beta^2 n_{\rm ex}(z,\beta)
\label{cbeta}
\end{equation}
\end{subequations}
Our numerical results coincide with Eqs.(\ref{nbeta}) and (\ref{cbeta}) remarkably well, with minor deviations for $|\delta z|\lesssim0.1${, as shown in Fig.~\ref{snr}}.

Applying Eq.(\ref{s}), we derive the thermodynamics of our model when $\alpha=0$.  
\begin{subequations}
Solving the saddle point of Eq.(\ref{part0}), we obtain the average energy density:
\be
\frac{1}{N_s}E(z,T)=\frac{r_0+n_{\rm ex}(z,T)}{2}\frac{T}{1+T}
\label{e0}
\ee
the specific heat:
\be
\frac{1}{N_s}C(z,T)=\frac{r_0+\frac{3}{2}n_{\rm ex}(z,T)}{2}\frac{1}{(1+T)^2}
\label{c0}
\ee
and the excess number density of redundant constraints:
\be
n_{\rm ex}(z,T)=n_0(z)\left(1+\frac{1}{T}\right)^{-1/2}
\label{n0}
\ee
\label{noweak}
\end{subequations}
where $r_0\equiv \frac{\delta z}{z}\Theta(\delta z)$. 

%

 As $n_0(z)$ is expected to be an analytic function of $z$, Eqs.(\ref{noweak}) indicate that $c$ converges to the one found in frozen network model in the limit $T\to0$:  $c=0$ when $\delta z<0$ and $c=\delta z/2z$ when $\delta z>0$ - the dashed yellow line in Fig.~\ref{theory}. This is our first central result,
which shows that our previous results hold even when the network is adaptive. 

Eqs.(\ref{noweak}) predict the energy, specific heat, and the number density of redundant constraints at an arbitrary temperature without any fitting parameter. The solid lines, shown in Fig.~\ref{thermo0}(a) and (b), are predictions of Eqs.(\ref{e0}) and (\ref{c0}), respectively, with $n_{\rm ex}$ as the numerical input. They are closely consistent with the data points, which confirms the annealed free energy approximation when $\alpha=0$. A $T^{1/2}$ power-law with numerical prefactor $n_0(z)=n_{\rm ex}(z,\infty)$ predicted by Eq.(\ref{n0}) coincides well with data points in Fig.~\ref{thermo0}(c).  

Extending to finite glass transition $T_g$ at $\alpha=0$, we find a correction vanishing as $\sqrt{\delta z}$ in addition to $c\approx \delta z/2z$, assuming $T_g\sim G\sim\delta z$ for $z>z_c$. But this correction is quantitatively unimportant as $n_0\leq0.03$ and does not change qualitatively the linear growth of the specific heat when $\delta z> 0$, as illustrated by the solid orange  line in Fig.~\ref{theory}. 

Our theoretic prediction that $n_{\rm ex}\to0$ when $T\to0$ validates the assumptions of \cite{Thorpe00,Chubynsky06,Barre09} that the energy of redundant bonds is proportional to their number, and that this number is  $R_0$ at $T=0$. 

\subsection{C. General case}
In the thermodynamic limit, $N_s\to\infty$, we take the saddle point of Eq.(\ref{part}),
\begin{subequations}
\be
\frac{2\partial s}{\partial n_r}=\ln\left(1+\frac{1}{T}\right)-\int\rd\omega D(\omega)\ln\left(1+\frac{1}{T}\frac{\alpha}{\omega^2+\alpha}\right)
\label{saddlen}
\ee
and
\be
\frac{2\delta s}{\delta D(\omega)}=(1-n_r)\ln\left(1+\frac{1}{T}\frac{\alpha}{\omega^2+\alpha}\right)
\label{saddleD}
\ee
\label{saddle}
\end{subequations}
and solve for energy,
\begin{multline}
\frac{1}{N_s}E(z,T,\alpha)=\frac{n_r(T)}{2}\frac{T}{1+{T}}\\+\frac{1-n_r(T)}{2}\int\rd\omega D(\omega,T)\frac{\alpha T}{\alpha+(\omega^2+\alpha){T}}
\label{e}
\end{multline}
The specific heat predictions from differentiating Eq.(\ref{e}) with numerical inputs $n_r(z,T,\alpha)$ and $D_{z,T,\alpha}(\omega)$ are plotted as solid lines in Fig.~\ref{cp}. (See Appendix Secs.~BCD for the temperature dependence of $D(\omega)$.) 
Notice that replacing $n_r(T)$ by $\delta z/z$ and $D(\omega, T)$ by its low-temperature limit $D(\omega)$ studied in \cite{Wyart08,During13,Yan13}, Eq.(\ref{e}) recovers exactly the one obtained in the frozen network model, whose predictions are plotted as dashed lines in Fig.~\ref{cp}. The dashed lines converge to the solid lines despite differences at high temperatures for weakly coordinated networks. 

In the limit $\alpha\to0$ and $T\ll\alpha$, Eq.(\ref{e}) converges to $E/N_s=T/2$, which indicates a constant specific heat $c=0.5$ when $\delta z<0$ independent of the models. This is shown by the solid orange line and the dashed yellow line in Fig.~\ref{theory}, and is our second key theoretical result showing the robustness of our conclusions for adaptive networks. 

\section{V. Conclusions}
In this work, we have studied the correlation between the elasticity of inherent structures and the thermodynamics in covalent glass-forming liquids using adaptive network models. We found numerically and explained theoretically why these models have a thermodynamic behavior similar to frozen network models \cite{Yan13} which captures nicely experimental facts. 

{ The main prediction conclusion of \cite{Yan13} is thus robust: as the coordination number approaches $z_c$ from above, elastic frustration vanishes. This leads both to an abundance of soft elastic modes, as well as a diminution of the number of directions in phase space that cost energy, which is directly proportional to the jump of specific heat. 
Below the rigidity transition, the elasticity of strong force network vanishes, thus the energy landscape is governed by the weak Van der Waals interactions. At these energy scale, all directions in contact space have a cost,  and thus the specific heat increases. Thus thermodynamic properties are governed by a critical point at $\delta z=0$, $\alpha=0$ where the jump of specific heat is zero. }
This prediction focuses on the configurational part of the jump of specific heat, since we considered only the energy minima in the metastable states.  In Appendix Sec.~E, we argue that the vibrational  contribution to this jump is so small in our models. { Thus the main prediction of the specific heat still holds, even when including the vibrational part}. 

{ Beyond network glasses, our main result potentially explains the correlation between   elasticity and the key aspects of the energy landscape in molecular glasses~\cite{Tatsumisago90,Bohmer92,Boolchand05}. Indeed  according to our work we expect glasses with a strong Boson peak to display less elastic frustration, so that they have a limited number of directions in phase space costing energy, see discussion in \cite{Yan13}. }

\begin{acknowledgements}
We thank E.~DeGiuli, G.~D\"uring, J.~Lin, E.~Lerner, C.~Sandford for discussions, and D.~Jacobs for sharing the pebble game code. This work has been supported primarily by the National Science Foundation Grant No. CBET-1236378, and partially by the Sloan
Fellowship, the NSF Grant No. DMR-1105387, and the Petroleum Research Fund Grant No. 52031-DNI9.
\end{acknowledgements}

\section{Appendix}
\appendix
\renewcommand{\theequation}{A\arabic{equation}}
\setcounter{equation}{0}
\setcounter{figure}{0}
\renewcommand{\thefigure}{A\arabic{figure}}
\subsection{A. Formalism of elastic energy}
\label{app_A}
The energy $H(\Gamma)$ of a given spring configuration $\Gamma\equiv\{\gamma\leftrightarrow\langle i,j\rangle\}$ is defined in Eq.(\ref{e1}) as a minimization on the position\textcolor{black}{s} of the nodes. This minimum can be calculated using conjugate gradient methods. However, for small mismatches $\epsilon$, it is more efficient to use  linear algebra~\cite{Yan13}, as we now recall.
Consider a displacement field $\delta\vec{R}_i\equiv\vec{R}_i-\vec{R}_{i0}$, where $\vec{R}_{i0}$ is the position of the node $i$ in the crystal described in the previous section. We define the distance $||\vec{R}_{i0}-\vec{R}_{j0}||\equiv r_{\langle i,j\rangle}$. At first order in $\delta\vec{R}_i$, the distance among  neighboring nodes can be written as:
\be
||\vec{R}_i-\vec{R}_j||=r_{\langle i,j\rangle}+\sum_{k}\ms_{\langle i,j\rangle,k}\delta\vec{R}_k+o(\delta\vec{R}^2)
\label{dist}
\ee
Where $\ms$ is the structure matrix, which gives the linear relation between displacements and changes of  distances, as indicated in Eq.(\ref{dist}). Minimizing Eq.(\ref{e1}), one gets:
\begin{multline}
H(\Gamma)=\min_{\{\delta\vec{R}_i\}}\left\{\frac{k}{2}\sum_{\gamma}(\sum_i\ms_{\gamma,i}\delta\vec{R}_i+\epsilon_{\gamma})^2\right.\\
\left.+\frac{k}{2}\sum_{\sigma} \frac{k_{\rm w}}{k}(\sum_{i}\ms_{\rm w{\ }\sigma,i}\delta\vec{R}_{i})^2+o(\delta\vec{R}^3)\right\}\\
=\min_{\{\delta\vec{R}_i\}}\frac{k}{2}\left[\langle\epsilon|\mpp|\epsilon\rangle+2\langle\epsilon|\ms|\delta\vec{R}\rangle+\langle\delta\vec{R}|\mm|\delta\vec{R}\rangle\right]
\label{linearexp}
\end{multline}
where we use bra-ket notations to indicate summation over edges or nodes, $\mpp$ projects the edge space to the subspace occupied by springs, $\mm\equiv\ms^t\ms+\frac{k_{\rm w}}{k}\ms_{\rm w}^t\ms_{\rm w}$ is the stiff matrix connecting the responding forces and displacements of nodes in an elastic network~\cite{Calladine78}, and $\bullet^t$ is our notation for the transpose of a matrix.
Solving Eq.(\ref{linearexp}), one finds the linear response,
\be
|\delta\vec{R}\rangle=-\mm^{-1}\ms^t|\epsilon\rangle
\label{response}
\ee
which for a given mismatch field $|\epsilon\rangle$ minimizes the elastic energy in Eq.(\ref{e1}). Inserting Eq.(\ref{response}) back into the linear approximation Eq.(\ref{linearexp}), we have~\cite{Yan13}:
\be
H(\Gamma)=\frac{k}{2}\langle\epsilon|\mpp-\ms\mm^{-1}\ms^t|\epsilon\rangle
=\frac{k}{2}\sum_{\Gamma}\epsilon_{\langle i,j\rangle}\mg_{\langle i,j\rangle,\langle l,m\rangle}\epsilon_{\langle l,m\rangle}
\label{Hamiltonian}
\ee
with $\mg=\mpp-\ms(\ms^t\ms+\frac{k_{\rw}}{k}\ms_{\rw}^t\ms_{\rw})^{-1}\ms^t$, and $\epsilon_{\langle i,j\rangle}=\epsilon_{\gamma}$ for $\Gamma=\{\gamma\leftrightarrow\langle i,j\rangle\}$.

\subsection{B. Density of states}
\label{app_B}
\begin{figure}[b]
\includegraphics[width=1.0\columnwidth]{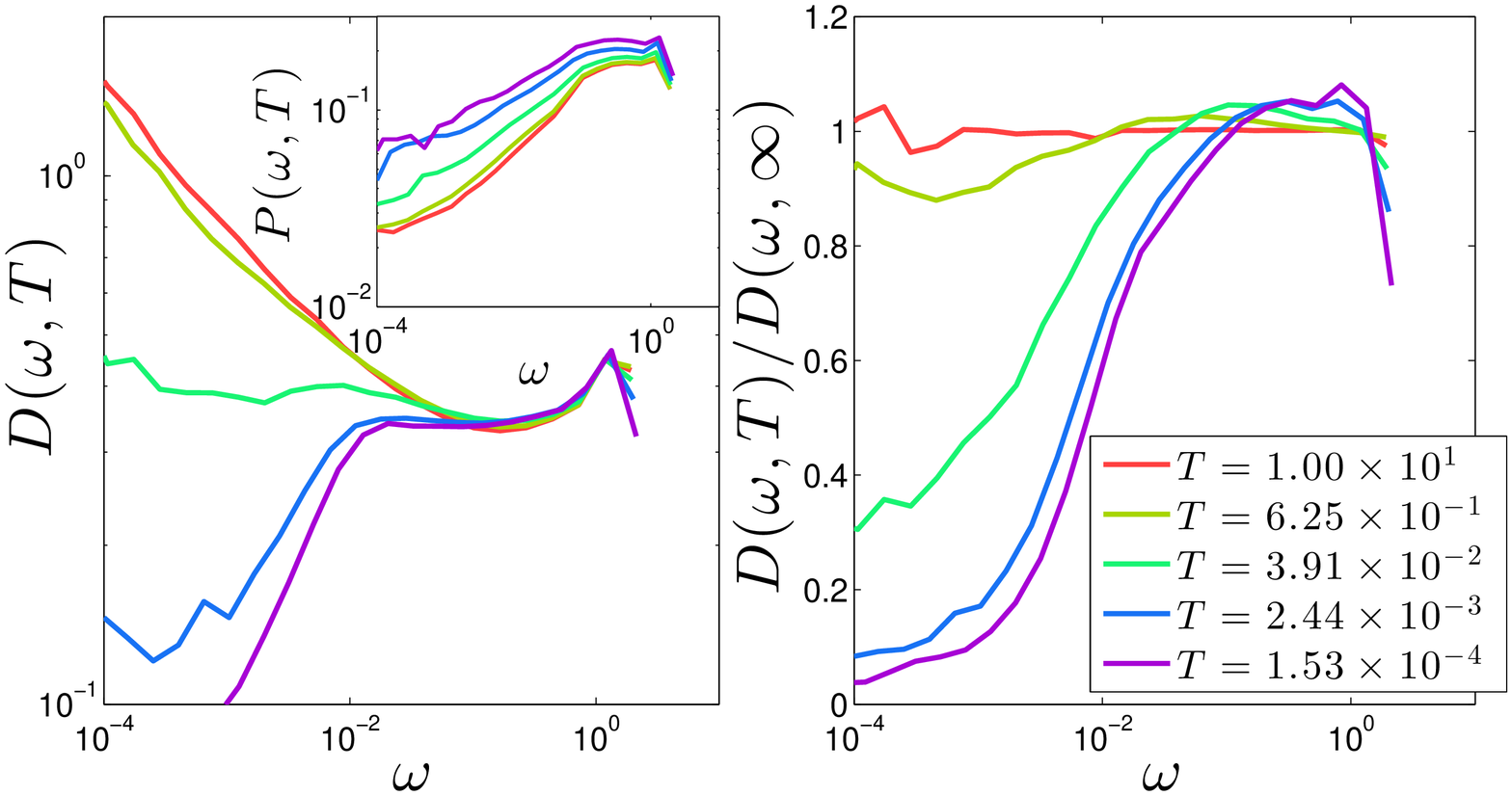}
\caption{\small{(Color online) Changes of density of states $D(\omega,T)$ with temperature for the same $z=-0.055$, $\alpha=0.0003$. Left: density of states in log-log scale. Right: density of states normalized by its $T=\infty$ value, emphasizing its difference under cooling. Inset: participation ratio $P(\omega,T)$ variation under cooling. 
}}\label{doscp}
\end{figure}

\begin{figure*}[ht]
\includegraphics[width=2.0\columnwidth]{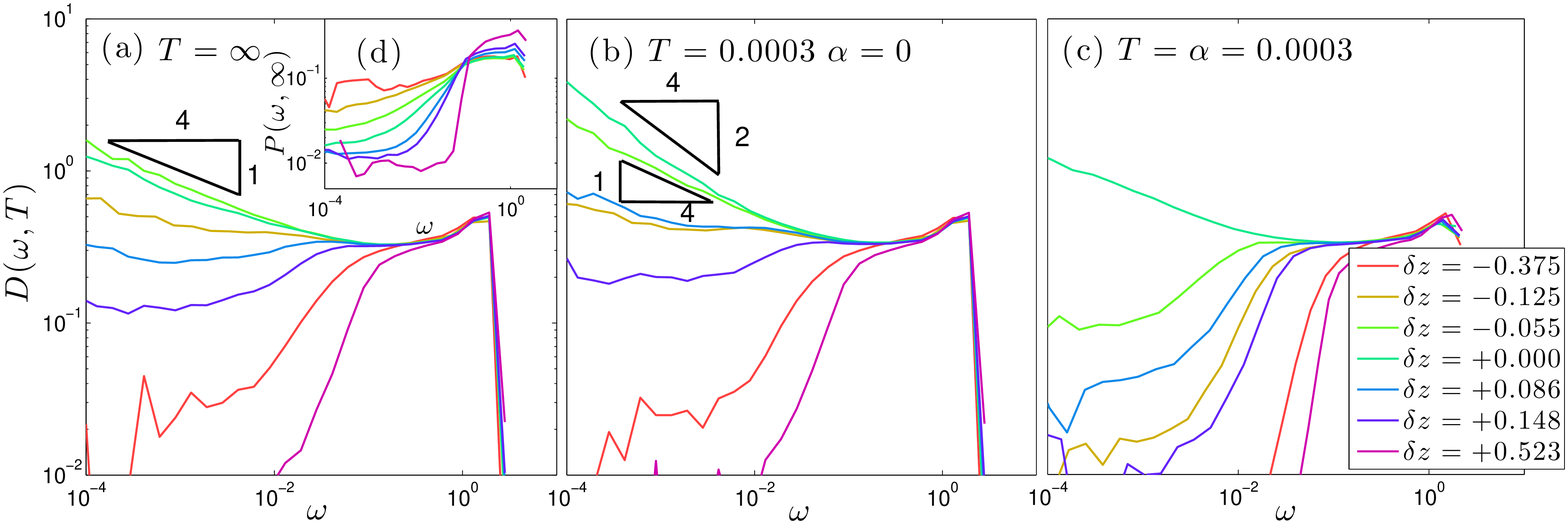}
\caption{\small{(Color online) Density of states $D(\omega,T)$ for adaptive networks with different $z$. (a) Random diluted networks $T=\infty$; a power law $D(\omega)\sim\omega^{-0.25}$ is shown in the low-frequency range of networks near $z_{cen}$. (b) Adaptive networks without weak constraints ($\alpha=0$) at $T=0.0003$; power laws with different exponents are shown for networks in the rigidity window: $D(\omega)\sim\omega^{-0.25}$ for $\delta z=-0.055$, $D(\omega)\sim\omega^{-0.5}$ for $\delta z=0.0$. (c) Adaptive networks with weak constraints ($\alpha=0.0003$) at $T\approx\alpha$; away from isostatic, the densities of states are gapped between zero frequency and Boson peak, where $D(\omega)\sim\omega^0$. Inset (d) is the participation ratio $P(\omega,T)$ at $T=\infty$, see text for definition.}}\label{dos}
\end{figure*}

We have shown the density of states converges to the one of mean-field networks~\cite{Yan14}. 
Cooling strongly suppresses low-frequency vibrational modes, as seen in Fig.~\ref{doscp}. 
This temperature effect on the density of states is primarily induced by the weak interactions: the density of states changes little under cooling when $\alpha=0$, as appeared in comparing (a) and (b) of Fig.~\ref{dos}. The slight change indicates that the density of states depends on the presence of redundant constraints. However, when $\alpha>0$, the low-temperature density of states strongly differs from its high-temperature counterpart, as shown in Fig.~\ref{dos}(a) and (c).

The modes that rarefy under cooling are localized vibrations. The participation ratio, $P(\omega)\equiv\frac{1}{Nd}(\sum_i\Psi_{\omega i}^2)^2/\sum_i\Psi_{\omega i}^4$, quantifies the extensity of characteristic modes: $P\to0$ corresponds to a localized mode, while $P\to1$ means that the mode extends over the system. Both the low and high-frequency ends of the density of states are reduced under cooling, but the modes in the middle are enhanced, as shown in the right panel of Fig.~\ref{doscp}. This agrees with the small participation ratio of modes with low and high frequencies, see Fig.~\ref{dos}(d). In fact, all modes become extended -- the participation ratio increases over the whole spectrum -- when the temperature decreases, as shown in the inset of Fig.~\ref{doscp}. 

In addition to localization, another prominent feature of reduced low-frequency modes is the power-law diverging density of states $D(\omega)\sim\omega^{\tilde{d}-1}$, see Fig.~\ref{dos}. The abundance of low-frequency localized modes appearing with a power law density of states signals the ``fractons'' that appear near the rigidity percolation~\cite{Alexander82,Feng84,Nakayama94}. 
The exponent of the diverging tail, in Fig.~\ref{dos}(a), implies the fracton dimension $\tilde{d}\approx0.75$, which is consistent with $0.78$ observed for the rigidity percolation~\cite{Feng85,Nakayama94}. Different fracton dimensions $\tilde{d}$ are observed for different coordination numbers in the case of rigidity window shown in Fig.~\ref{dos}(b), although more work would be needed to establish this fact empirically. 

We discuss when the temperature affects the mode with frequency $\omega$ in Appendix Sec. C and show illustrations of ``fractons'' in Appendix Sec. D.


\subsection{C. Adaptation effects on density of states}
\label{app_C}
When $\alpha>0$, following Eq.(\ref{coupmat}), we find out the typical elastic energy corresponding to a mode of frequency $\omega$ scales as ${\alpha}/({\omega^2+\alpha})$, which is proportional to $\alpha$ for $\omega\sim1$, while proportional to $1$ when $\omega\ll\sqrt{\alpha}$. This implies that the elastic energy in the degrees of freedom corresponding to the modes of low-frequency is of the same magnitude as the one in the redundant constraints. Similar to the redundant constraints, these low-frequency modes are reduced under cooling. 

From Eq.(\ref{saddleD}), 
$T^*(\omega,\alpha)\sim{\alpha}/({\omega^2+\alpha})$ gives an estimate of the temperature scale the mode $\omega$ begins to be reduced. The adaptation effect at this temperature scale can be seen in the right panel of Fig.~\ref{doscp}. For example, the dashed green line at $T\approx0.04\ll1$ shows a density of states with frequencies $\omega\lesssim\sqrt{\alpha}\approx0.01$ strongly suppressed, while the shape of the density of states with $\omega\approx0.1$ and above is almost unchanged. The dotted purple line, $T\approx10^{-4}\sim\alpha$, shows a density of states whose highest frequency $\omega\sim1$ is also significantly reduced. 

\subsection{D. Fractons}
\label{app_D}
\begin{figure*}[ht]
\includegraphics[width=2.0\columnwidth]{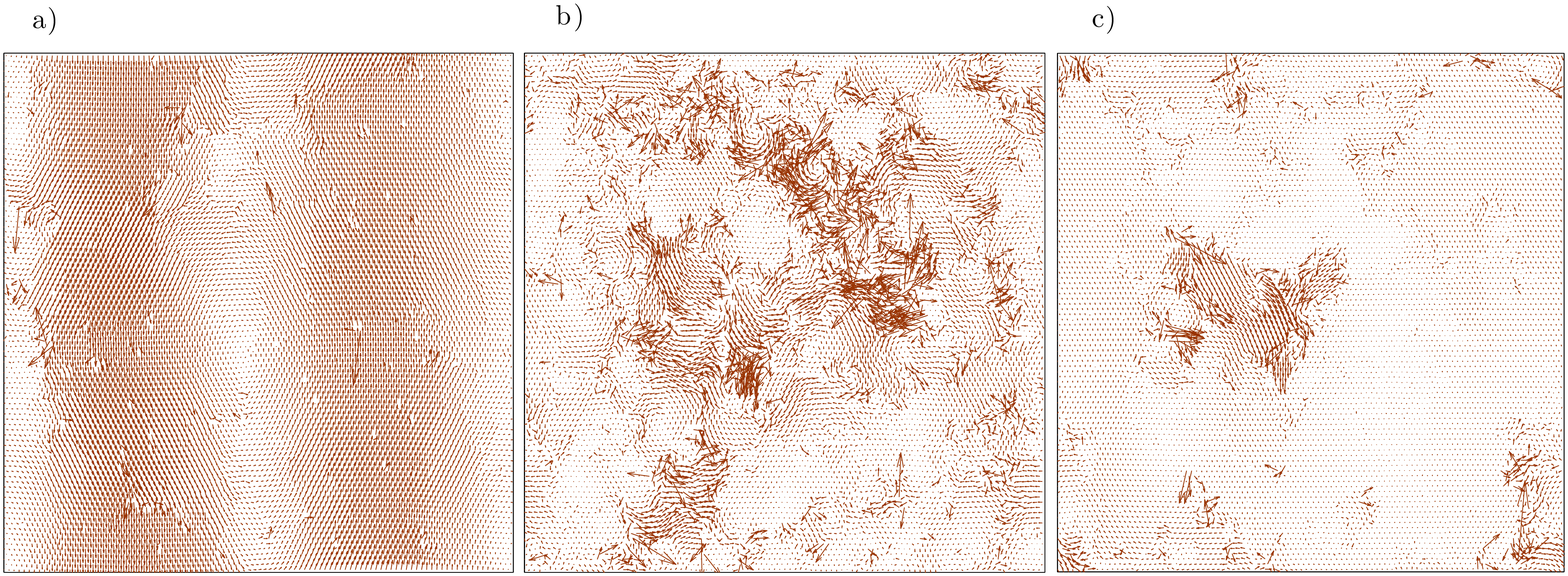}
\caption{\small{(Color online) Vector plots of vibrational modes in randomly diluted networks, $N=100\times100$. (a) A typical Debye mode, $\delta z=0.501$, $\omega=0.017$. (b) A typical anomalous mode on boson peak, $\delta z=-0.049$, $\omega=0.011$. (c) A typical fracton, $\delta z=-0.049$, $\omega=0.0007$.}}\label{mode}
\end{figure*}

\begin{figure}[b]
\centering
\includegraphics[width=.9\columnwidth]{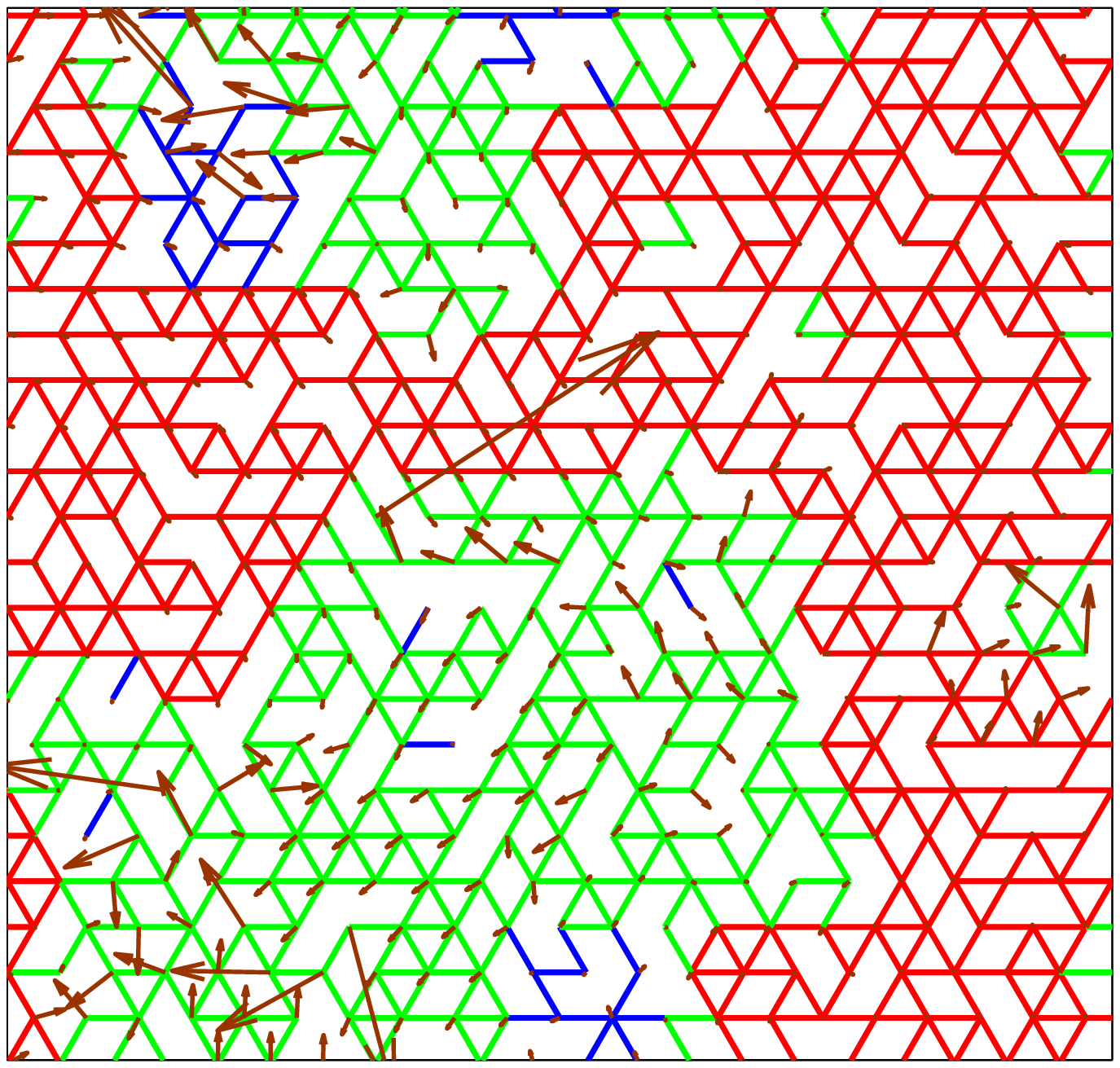}
\caption{\small{(Color online) Correlation between a low-frequency fractal mode and isostatic clusters. A network configuration ($\delta z=-0.042$) is shown with its springs in the over-constrained regions colored in red, in the isostatic regions colored in green, and in the floppy regions colored in blue. A typical fracton ($\omega=5\times10^{-4}$) specified in this configuration is plotted on top.
}}\label{coupl}
\end{figure}

``Fractons'' are different from either the low-frequency Debye modes or the anomalous modes on the boson peak, as shown in Fig.~\ref{mode}. They (Fig.~\ref{mode}(c)) are localized and random compared to the Debye modes (Fig.~\ref{mode}(a)), and concentrated on the fractal sets with sharp boundaries, unlike the extended anomalous modes (Fig.~\ref{mode}(b)). 
The ``fractons'' are associated with the collective motion of large isostatic or nearly isostatic regions as shown in Fig.~\ref{coupl}. 

\subsection{E. Vibrational entropy contribution}
\label{app_E}
The structure the elastic potential evolve with temperature in the liquid phase of the adaptive network model. Freezing into a glass phase eliminates this variability and leads to a contribution to the jump of specific heat~\cite{Wyart10}. Our model currently ignores the vibrational part of the specific heat, which incorporates that the shape of the inherent structure evolves with temperature - not only its bottom energy. We estimate this contribution from vibrations in this subsection and argue that is is not significant for the models we consider.

The vibrational entropy includes both linear $\omega>0$ and floppy $\omega=0$ vibration modes~\cite{Wyart10}:
\be
s_{vib}(T)=[1-n_r(T)]\int\rd\omega D(\omega,T)\ln\frac{eT}{\hbar\omega}+f(T)\ln\Lambda
\label{svib}
\ee
$\Lambda$ sets a cutoff volume for floppy modes, which is approximately the atomic spacing measured in the Lindemann's length: $\Lambda\approx(1/0.15)^d$~\cite{Nelson02}, of order $10^3$ in 3D~\cite{Lindemann10}. $f$ is the floppy mode density, dual to the number density of redundant constraints  $f(T)=-\delta z/z+n_r(T)$ and thus $\partial f(T)/\partial T=\partial n_r(T)/\partial T$. The jump of specific heat follows:
\begin{widetext}
\be
\Delta c_{vib}=\left.T_g\frac{\partial n_r(T)}{\partial T}\right|_{T_g}\left[\ln\Lambda-\int\rd\omega D_{T_g}(\omega)\ln\frac{eT_g}{\hbar\omega}\right]+[1-n_r(T_g)]\int\rd\omega\left.T_g\frac{\partial D_T(\omega)}{\partial T}\right|_{T_g}\ln\frac{eT_g}{\hbar\omega}
\label{cvib}
\ee
\end{widetext}
The derivatives on $\ln T$ in Eq.(\ref{svib}), continuous at the glass transition, have been subtracted. 

We estimate the upper limit of the vibrational contribution. 
(1) The first term in Eq.(\ref{cvib}): Debye frequency $\omega_D$ sets the upper limit of the integral in the bracket, $-\ln({eT_g}/{\hbar\omega_D})$. As the glass transition temperature $T_g$ and Debye temperature $\theta_D=\hbar\omega_D/k_B$ are usually of the same order, the bracket in the first term is dominated by $\ln\Lambda$. 
From Eqs.(11), we have ${\partial n_r}/{\partial\ln T}|_{T_g}\approx\frac{1}{2}n_{ex}(T_g)\lesssim\frac{1}{2}n_0\sqrt{T_g}\lesssim0.02\sqrt{\alpha}$, and $\ln\Lambda\approx5$ in 2D. Compared to the specific heat values, which are of order one shown in Fig.~7, and the scalings of the minima $-0.1/\ln\alpha$ given in \cite{Yan13}, the contribution, $0.1\sqrt{\alpha}$, is insignificant if $0<\alpha<0.1$. 

(2) The second term in Eq.(\ref{cvib}): The upper limit of the bracket is $1$. 
Replacing $\ln({eT}/{\hbar\omega})$ with its upper limit $\ln\Lambda$, we simplify the integral to $\int\rd\omega T{\partial D}/{\partial T}$. 
We can estimate the upper limit of the derivative in the integral approximately by ${\Delta n_T}/{\Delta\ln T}$, where $\Delta n_T$ is the number density of the modes reduced under cooling. $\Delta n_T\approx0.2\int_0^{0.01}\omega^{-0.25}\rd\omega\approx0.01$, roughly the number fraction of ``fractons'' suppressed under cooling. Together, the upper limit of the contribution of the second term is $\Delta n_T/\ln10\times\ln\Lambda\approx0.03$, which is moderate compared to the values of order one. 

Therefore, the vibrational entropy contributes mildly to the jump of specific heat and does not change the qualitative behavior of $\Delta c$ in our model of network glasses. 

\bibliography{Wyartbibnew}

\begin{thebibliography}{68}%
\makeatletter
\providecommand \@ifxundefined [1]{%
 \@ifx{#1\undefined}
}%
\providecommand \@ifnum [1]{%
 \ifnum #1\expandafter \@firstoftwo
 \else \expandafter \@secondoftwo
 \fi
}%
\providecommand \@ifx [1]{%
 \ifx #1\expandafter \@firstoftwo
 \else \expandafter \@secondoftwo
 \fi
}%
\providecommand \natexlab [1]{#1}%
\providecommand \enquote  [1]{``#1''}%
\providecommand \bibnamefont  [1]{#1}%
\providecommand \bibfnamefont [1]{#1}%
\providecommand \citenamefont [1]{#1}%
\providecommand \href@noop [0]{\@secondoftwo}%
\providecommand \href [0]{\begingroup \@sanitize@url \@href}%
\providecommand \@href[1]{\@@startlink{#1}\@@href}%
\providecommand \@@href[1]{\endgroup#1\@@endlink}%
\providecommand \@sanitize@url [0]{\catcode `\\12\catcode `\$12\catcode
  `\&12\catcode `\#12\catcode `\^12\catcode `\_12\catcode `\%12\relax}%
\providecommand \@@startlink[1]{}%
\providecommand \@@endlink[0]{}%
\providecommand \url  [0]{\begingroup\@sanitize@url \@url }%
\providecommand \@url [1]{\endgroup\@href {#1}{\urlprefix }}%
\providecommand \urlprefix  [0]{URL }%
\providecommand \Eprint [0]{\href }%
\providecommand \doibase [0]{http://dx.doi.org/}%
\providecommand \selectlanguage [0]{\@gobble}%
\providecommand \bibinfo  [0]{\@secondoftwo}%
\providecommand \bibfield  [0]{\@secondoftwo}%
\providecommand \translation [1]{[#1]}%
\providecommand \BibitemOpen [0]{}%
\providecommand \bibitemStop [0]{}%
\providecommand \bibitemNoStop [0]{.\EOS\space}%
\providecommand \EOS [0]{\spacefactor3000\relax}%
\providecommand \BibitemShut  [1]{\csname bibitem#1\endcsname}%
\let\auto@bib@innerbib\@empty
\bibitem [{\citenamefont {Debenedetti}\ and\ \citenamefont
  {Stillinger}(2001)}]{Debenedetti01}%
  \BibitemOpen
  \bibfield  {author} {\bibinfo {author} {\bibfnamefont {P.~G.}\ \bibnamefont
  {Debenedetti}}\ and\ \bibinfo {author} {\bibfnamefont {F.~H.}\ \bibnamefont
  {Stillinger}},\ }\href {http://dx.doi.org/10.1038/35065704} {\bibfield
  {journal} {\bibinfo  {journal} {Nature}\ }\textbf {\bibinfo {volume} {410}},\
  \bibinfo {pages} {259} (\bibinfo {year} {2001})}\BibitemShut {NoStop}%
\bibitem [{\citenamefont {Stillinger}\ and\ \citenamefont
  {Weber}(1984)}]{Stillinger84}%
  \BibitemOpen
  \bibfield  {author} {\bibinfo {author} {\bibfnamefont {F.~H.}\ \bibnamefont
  {Stillinger}}\ and\ \bibinfo {author} {\bibfnamefont {T.~A.}\ \bibnamefont
  {Weber}},\ }\href@noop {} {\bibfield  {journal} {\bibinfo  {journal}
  {Science(Washington, DC)}\ }\textbf {\bibinfo {volume} {225}},\ \bibinfo
  {pages} {983} (\bibinfo {year} {1984})}\BibitemShut {NoStop}%
\bibitem [{\citenamefont {Adam}\ and\ \citenamefont {Gibbs}(1965)}]{Adam65}%
  \BibitemOpen
  \bibfield  {author} {\bibinfo {author} {\bibfnamefont {G.}~\bibnamefont
  {Adam}}\ and\ \bibinfo {author} {\bibfnamefont {J.~H.}\ \bibnamefont
  {Gibbs}},\ }\href {\doibase http://dx.doi.org/10.1063/1.1696442} {\bibfield
  {journal} {\bibinfo  {journal} {The Journal of Chemical Physics}\ }\textbf
  {\bibinfo {volume} {43}},\ \bibinfo {pages} {139} (\bibinfo {year}
  {1965})}\BibitemShut {NoStop}%
\bibitem [{\citenamefont {Kirkpatrick}\ \emph {et~al.}(1989)\citenamefont
  {Kirkpatrick}, \citenamefont {Thirumalai},\ and\ \citenamefont
  {Wolynes}}]{Kirkpatrick89}%
  \BibitemOpen
  \bibfield  {author} {\bibinfo {author} {\bibfnamefont {T.~R.}\ \bibnamefont
  {Kirkpatrick}}, \bibinfo {author} {\bibfnamefont {D.}~\bibnamefont
  {Thirumalai}}, \ and\ \bibinfo {author} {\bibfnamefont {P.~G.}\ \bibnamefont
  {Wolynes}},\ }\href {\doibase 10.1103/PhysRevA.40.1045} {\bibfield  {journal}
  {\bibinfo  {journal} {Phys. Rev. A}\ }\textbf {\bibinfo {volume} {40}},\
  \bibinfo {pages} {1045} (\bibinfo {year} {1989})}\BibitemShut {NoStop}%
\bibitem [{\citenamefont {Lubchenko}\ and\ \citenamefont
  {Wolynes}(2007)}]{Lubchenko07}%
  \BibitemOpen
  \bibfield  {author} {\bibinfo {author} {\bibfnamefont {V.}~\bibnamefont
  {Lubchenko}}\ and\ \bibinfo {author} {\bibfnamefont {P.~G.}\ \bibnamefont
  {Wolynes}},\ }\href {\doibase 10.1146/annurev.physchem.58.032806.104653}
  {\bibfield  {journal} {\bibinfo  {journal} {Annual Review of Physical
  Chemistry}\ }\textbf {\bibinfo {volume} {58}},\ \bibinfo {pages} {235}
  (\bibinfo {year} {2007})}\BibitemShut {NoStop}%
\bibitem [{\citenamefont {Bouchaud}\ and\ \citenamefont
  {Biroli}(2004)}]{Bouchaud04}%
  \BibitemOpen
  \bibfield  {author} {\bibinfo {author} {\bibfnamefont {J.-P.}\ \bibnamefont
  {Bouchaud}}\ and\ \bibinfo {author} {\bibfnamefont {G.}~\bibnamefont
  {Biroli}},\ }\href {\doibase http://dx.doi.org/10.1063/1.1796231} {\bibfield
  {journal} {\bibinfo  {journal} {The Journal of Chemical Physics}\ }\textbf
  {\bibinfo {volume} {121}},\ \bibinfo {pages} {7347} (\bibinfo {year}
  {2004})}\BibitemShut {NoStop}%
\bibitem [{\citenamefont {Chamberlin}(1999)}]{Chamberlin99}%
  \BibitemOpen
  \bibfield  {author} {\bibinfo {author} {\bibfnamefont {R.~V.}\ \bibnamefont
  {Chamberlin}},\ }\href {\doibase 10.1103/PhysRevLett.82.2520} {\bibfield
  {journal} {\bibinfo  {journal} {Phys. Rev. Lett.}\ }\textbf {\bibinfo
  {volume} {82}},\ \bibinfo {pages} {2520} (\bibinfo {year}
  {1999})}\BibitemShut {NoStop}%
\bibitem [{\citenamefont {Dyre}(2006)}]{Dyre06}%
  \BibitemOpen
  \bibfield  {author} {\bibinfo {author} {\bibfnamefont {J.~C.}\ \bibnamefont
  {Dyre}},\ }\href@noop {} {\bibfield  {journal} {\bibinfo  {journal} {Reviews
  of modern physics}\ }\textbf {\bibinfo {volume} {78}},\ \bibinfo {pages}
  {953} (\bibinfo {year} {2006})}\BibitemShut {NoStop}%
\bibitem [{\citenamefont {Chandler}\ and\ \citenamefont
  {Garrahan}(2010)}]{Chandler10}%
  \BibitemOpen
  \bibfield  {author} {\bibinfo {author} {\bibfnamefont {D.}~\bibnamefont
  {Chandler}}\ and\ \bibinfo {author} {\bibfnamefont {J.~P.}\ \bibnamefont
  {Garrahan}},\ }\href {\doibase 10.1146/annurev.physchem.040808.090405}
  {\bibfield  {journal} {\bibinfo  {journal} {Annual Review of Physical
  Chemistry}\ }\textbf {\bibinfo {volume} {61}},\ \bibinfo {pages} {191}
  (\bibinfo {year} {2010})},\ \bibinfo {note} {pMID: 20055676},\ \Eprint
  {http://arxiv.org/abs/http://dx.doi.org/10.1146/annurev.physchem.040808.090405}
  {http://dx.doi.org/10.1146/annurev.physchem.040808.090405} \BibitemShut
  {NoStop}%
\bibitem [{\citenamefont {Martinez}\ and\ \citenamefont
  {Angell}(2001)}]{Martinez01}%
  \BibitemOpen
  \bibfield  {author} {\bibinfo {author} {\bibfnamefont {L.-M.}\ \bibnamefont
  {Martinez}}\ and\ \bibinfo {author} {\bibfnamefont {C.~A.}\ \bibnamefont
  {Angell}},\ }\href {\doibase http://dx.doi.org/10.1038/35070517} {\bibfield
  {journal} {\bibinfo  {journal} {Nature}\ }\textbf {\bibinfo {volume} {410}},\
  \bibinfo {pages} {663} (\bibinfo {year} {2001})}\BibitemShut {NoStop}%
\bibitem [{\citenamefont {Wang}\ \emph {et~al.}(2006)\citenamefont {Wang},
  \citenamefont {Angell},\ and\ \citenamefont {Richert}}]{Wang06}%
  \BibitemOpen
  \bibfield  {author} {\bibinfo {author} {\bibfnamefont {L.-M.}\ \bibnamefont
  {Wang}}, \bibinfo {author} {\bibfnamefont {C.~A.}\ \bibnamefont {Angell}}, \
  and\ \bibinfo {author} {\bibfnamefont {R.}~\bibnamefont {Richert}},\ }\href
  {\doibase http://dx.doi.org/10.1063/1.2244551} {\bibfield  {journal}
  {\bibinfo  {journal} {The Journal of Chemical Physics}\ }\textbf {\bibinfo
  {volume} {125}},\ \bibinfo {eid} {074505} (\bibinfo {year}
  {2006})}\BibitemShut {NoStop}%
\bibitem [{\citenamefont {Hall}\ and\ \citenamefont {Wolynes}(2003)}]{Hall03}%
  \BibitemOpen
  \bibfield  {author} {\bibinfo {author} {\bibfnamefont {R.~W.}\ \bibnamefont
  {Hall}}\ and\ \bibinfo {author} {\bibfnamefont {P.~G.}\ \bibnamefont
  {Wolynes}},\ }\href {\doibase 10.1103/PhysRevLett.90.085505} {\bibfield
  {journal} {\bibinfo  {journal} {Phys. Rev. Lett.}\ }\textbf {\bibinfo
  {volume} {90}},\ \bibinfo {pages} {085505} (\bibinfo {year}
  {2003})}\BibitemShut {NoStop}%
\bibitem [{\citenamefont {Bevzenko}\ and\ \citenamefont
  {Lubchenko}(2009)}]{Bevzenko09}%
  \BibitemOpen
  \bibfield  {author} {\bibinfo {author} {\bibfnamefont {D.}~\bibnamefont
  {Bevzenko}}\ and\ \bibinfo {author} {\bibfnamefont {V.}~\bibnamefont
  {Lubchenko}},\ }\href@noop {} {\bibfield  {journal} {\bibinfo  {journal} {The
  Journal of Physical Chemistry B}\ }\textbf {\bibinfo {volume} {113}},\
  \bibinfo {pages} {16337} (\bibinfo {year} {2009})}\BibitemShut {NoStop}%
\bibitem [{\citenamefont {de~Souza}\ and\ \citenamefont
  {Harrowell}(2009{\natexlab{a}})}]{Souza09}%
  \BibitemOpen
  \bibfield  {author} {\bibinfo {author} {\bibfnamefont {V.~K.}\ \bibnamefont
  {de~Souza}}\ and\ \bibinfo {author} {\bibfnamefont {P.}~\bibnamefont
  {Harrowell}},\ }\href@noop {} {\bibfield  {journal} {\bibinfo  {journal}
  {Proceedings of the National Academy of Sciences}\ }\textbf {\bibinfo
  {volume} {106}},\ \bibinfo {pages} {15136} (\bibinfo {year}
  {2009}{\natexlab{a}})}\BibitemShut {NoStop}%
\bibitem [{\citenamefont {Rabochiy}\ and\ \citenamefont
  {Lubchenko}(2013)}]{Rabochiy13}%
  \BibitemOpen
  \bibfield  {author} {\bibinfo {author} {\bibfnamefont {P.}~\bibnamefont
  {Rabochiy}}\ and\ \bibinfo {author} {\bibfnamefont {V.}~\bibnamefont
  {Lubchenko}},\ }\href@noop {} {\bibfield  {journal} {\bibinfo  {journal} {The
  Journal of chemical physics}\ }\textbf {\bibinfo {volume} {138}},\ \bibinfo
  {pages} {12A534} (\bibinfo {year} {2013})}\BibitemShut {NoStop}%
\bibitem [{\citenamefont {Anderson}(1981)}]{Phillips81}%
  \BibitemOpen
  \bibfield  {author} {\bibinfo {author} {\bibfnamefont {A.}~\bibnamefont
  {Anderson}},\ }\href@noop {} {\emph {\bibinfo {title} {Amorphous Solids: Low
  Temperature Properties}}},\ edited by\ \bibinfo {editor} {\bibfnamefont
  {W.~A.}\ \bibnamefont {Phillips}},\ \bibinfo {series} {Topics in Current
  Physics}, Vol.~\bibinfo {volume} {24}\ (\bibinfo  {publisher} {Springer,
  Berlin},\ \bibinfo {year} {1981})\BibitemShut {NoStop}%
\bibitem [{\citenamefont {Ngai}\ \emph {et~al.}(1997)\citenamefont {Ngai},
  \citenamefont {Sokolov},\ and\ \citenamefont {Steffen}}]{Ngai97}%
  \BibitemOpen
  \bibfield  {author} {\bibinfo {author} {\bibfnamefont {K.}~\bibnamefont
  {Ngai}}, \bibinfo {author} {\bibfnamefont {A.}~\bibnamefont {Sokolov}}, \
  and\ \bibinfo {author} {\bibfnamefont {W.}~\bibnamefont {Steffen}},\
  }\href@noop {} {\bibfield  {journal} {\bibinfo  {journal} {Journal of
  Chemical Physics}\ }\textbf {\bibinfo {volume} {107}},\ \bibinfo {pages}
  {5268} (\bibinfo {year} {1997})}\BibitemShut {NoStop}%
\bibitem [{\citenamefont {Novikov}\ \emph {et~al.}(2005)\citenamefont
  {Novikov}, \citenamefont {Ding},\ and\ \citenamefont {Sokolov}}]{Novikov05}%
  \BibitemOpen
  \bibfield  {author} {\bibinfo {author} {\bibfnamefont {V.~N.}\ \bibnamefont
  {Novikov}}, \bibinfo {author} {\bibfnamefont {Y.}~\bibnamefont {Ding}}, \
  and\ \bibinfo {author} {\bibfnamefont {A.~P.}\ \bibnamefont {Sokolov}},\
  }\href@noop {} {\bibfield  {journal} {\bibinfo  {journal} {Phys. Rev. E}\
  }\textbf {\bibinfo {volume} {71}},\ \bibinfo {pages} {061501} (\bibinfo
  {year} {2005})}\BibitemShut {NoStop}%
\bibitem [{\citenamefont {Tatsumisago}\ \emph {et~al.}(1990)\citenamefont
  {Tatsumisago}, \citenamefont {Halfpap}, \citenamefont {Green}, \citenamefont
  {Lindsay},\ and\ \citenamefont {Angell}}]{Tatsumisago90}%
  \BibitemOpen
  \bibfield  {author} {\bibinfo {author} {\bibfnamefont {M.}~\bibnamefont
  {Tatsumisago}}, \bibinfo {author} {\bibfnamefont {B.~L.}\ \bibnamefont
  {Halfpap}}, \bibinfo {author} {\bibfnamefont {J.~L.}\ \bibnamefont {Green}},
  \bibinfo {author} {\bibfnamefont {S.~M.}\ \bibnamefont {Lindsay}}, \ and\
  \bibinfo {author} {\bibfnamefont {C.~A.}\ \bibnamefont {Angell}},\ }\href
  {\doibase 10.1103/PhysRevLett.64.1549} {\bibfield  {journal} {\bibinfo
  {journal} {Phys. Rev. Lett.}\ }\textbf {\bibinfo {volume} {64}},\ \bibinfo
  {pages} {1549} (\bibinfo {year} {1990})}\BibitemShut {NoStop}%
\bibitem [{\citenamefont {Kamitakahara}\ \emph {et~al.}(1991)\citenamefont
  {Kamitakahara}, \citenamefont {Cappelletti}, \citenamefont {Boolchand},
  \citenamefont {Halfpap}, \citenamefont {Gompf}, \citenamefont {Neumann},\
  and\ \citenamefont {Mutka}}]{Kamitakahara91}%
  \BibitemOpen
  \bibfield  {author} {\bibinfo {author} {\bibfnamefont {W.~A.}\ \bibnamefont
  {Kamitakahara}}, \bibinfo {author} {\bibfnamefont {R.~L.}\ \bibnamefont
  {Cappelletti}}, \bibinfo {author} {\bibfnamefont {P.}~\bibnamefont
  {Boolchand}}, \bibinfo {author} {\bibfnamefont {B.}~\bibnamefont {Halfpap}},
  \bibinfo {author} {\bibfnamefont {F.}~\bibnamefont {Gompf}}, \bibinfo
  {author} {\bibfnamefont {D.~A.}\ \bibnamefont {Neumann}}, \ and\ \bibinfo
  {author} {\bibfnamefont {H.}~\bibnamefont {Mutka}},\ }\href {\doibase
  10.1103/PhysRevB.44.94} {\bibfield  {journal} {\bibinfo  {journal} {Phys.
  Rev. B}\ }\textbf {\bibinfo {volume} {44}},\ \bibinfo {pages} {94} (\bibinfo
  {year} {1991})}\BibitemShut {NoStop}%
\bibitem [{\citenamefont {Selvanathan}\ \emph {et~al.}(1999)\citenamefont
  {Selvanathan}, \citenamefont {Bresser}, \citenamefont {Boolchand},\ and\
  \citenamefont {Goodman}}]{Selvanathan99}%
  \BibitemOpen
  \bibfield  {author} {\bibinfo {author} {\bibfnamefont {D.}~\bibnamefont
  {Selvanathan}}, \bibinfo {author} {\bibfnamefont {W.}~\bibnamefont
  {Bresser}}, \bibinfo {author} {\bibfnamefont {P.}~\bibnamefont {Boolchand}},
  \ and\ \bibinfo {author} {\bibfnamefont {B.}~\bibnamefont {Goodman}},\ }\href
  {\doibase http://dx.doi.org/10.1016/S0038-1098(99)00248-3} {\bibfield
  {journal} {\bibinfo  {journal} {Solid State Communications}\ }\textbf
  {\bibinfo {volume} {111}},\ \bibinfo {pages} {619 } (\bibinfo {year}
  {1999})}\BibitemShut {NoStop}%
\bibitem [{\citenamefont {Maxwell}(1864)}]{Maxwell64}%
  \BibitemOpen
  \bibfield  {author} {\bibinfo {author} {\bibfnamefont {J.}~\bibnamefont
  {Maxwell}},\ }\href@noop {} {\bibfield  {journal} {\bibinfo  {journal}
  {Philos. Mag.}\ }\textbf {\bibinfo {volume} {27}},\ \bibinfo {pages} {294}
  (\bibinfo {year} {1864})}\BibitemShut {NoStop}%
\bibitem [{\citenamefont {Phillips}(1979)}]{Phillips79}%
  \BibitemOpen
  \bibfield  {author} {\bibinfo {author} {\bibfnamefont {J.}~\bibnamefont
  {Phillips}},\ }\href {\doibase
  http://dx.doi.org/10.1016/0022-3093(79)90033-4} {\bibfield  {journal}
  {\bibinfo  {journal} {Journal of Non-Crystalline Solids}\ }\textbf {\bibinfo
  {volume} {34}},\ \bibinfo {pages} {153 } (\bibinfo {year}
  {1979})}\BibitemShut {NoStop}%
\bibitem [{\citenamefont {{Phillips}}\ and\ \citenamefont
  {{Thorpe}}(1985)}]{Phillips85}%
  \BibitemOpen
  \bibfield  {author} {\bibinfo {author} {\bibfnamefont {J.~C.}\ \bibnamefont
  {{Phillips}}}\ and\ \bibinfo {author} {\bibfnamefont {M.~F.}\ \bibnamefont
  {{Thorpe}}},\ }\href@noop {} {\bibfield  {journal} {\bibinfo  {journal}
  {Sol.\ State Comm.}\ }\textbf {\bibinfo {volume} {53}},\ \bibinfo {pages}
  {699} (\bibinfo {year} {1985})}\BibitemShut {NoStop}%
\bibitem [{\citenamefont {B\"ohmer}\ and\ \citenamefont
  {Angell}(1992)}]{Bohmer92}%
  \BibitemOpen
  \bibfield  {author} {\bibinfo {author} {\bibfnamefont {R.}~\bibnamefont
  {B\"ohmer}}\ and\ \bibinfo {author} {\bibfnamefont {C.~A.}\ \bibnamefont
  {Angell}},\ }\href {\doibase 10.1103/PhysRevB.45.10091} {\bibfield  {journal}
  {\bibinfo  {journal} {Phys. Rev. B}\ }\textbf {\bibinfo {volume} {45}},\
  \bibinfo {pages} {10091} (\bibinfo {year} {1992})}\BibitemShut {NoStop}%
\bibitem [{\citenamefont {Micoulaut}\ and\ \citenamefont
  {Boolchand}(2003)}]{Micoulaut03a}%
  \BibitemOpen
  \bibfield  {author} {\bibinfo {author} {\bibfnamefont {M.}~\bibnamefont
  {Micoulaut}}\ and\ \bibinfo {author} {\bibfnamefont {P.}~\bibnamefont
  {Boolchand}},\ }\href {\doibase 10.1103/PhysRevLett.91.159601} {\bibfield
  {journal} {\bibinfo  {journal} {Phys. Rev. Lett.}\ }\textbf {\bibinfo
  {volume} {91}},\ \bibinfo {pages} {159601} (\bibinfo {year}
  {2003})}\BibitemShut {NoStop}%
\bibitem [{\citenamefont {Trachenko}\ \emph {et~al.}(2000)\citenamefont
  {Trachenko}, \citenamefont {Dove}, \citenamefont {Harris},\ and\
  \citenamefont {Heine}}]{Trachenko00}%
  \BibitemOpen
  \bibfield  {author} {\bibinfo {author} {\bibfnamefont {K.~O.}\ \bibnamefont
  {Trachenko}}, \bibinfo {author} {\bibfnamefont {M.~T.}\ \bibnamefont {Dove}},
  \bibinfo {author} {\bibfnamefont {M.~J.}\ \bibnamefont {Harris}}, \ and\
  \bibinfo {author} {\bibfnamefont {V.}~\bibnamefont {Heine}},\ }\href@noop {}
  {\bibfield  {journal} {\bibinfo  {journal} {Journal of Physics: Condensed
  Matter}\ }\textbf {\bibinfo {volume} {12}},\ \bibinfo {pages} {8041}
  (\bibinfo {year} {2000})}\BibitemShut {NoStop}%
\bibitem [{\citenamefont {O'Hern}\ \emph {et~al.}(2003)\citenamefont {O'Hern},
  \citenamefont {Silbert}, \citenamefont {Liu},\ and\ \citenamefont
  {Nagel}}]{OHern03}%
  \BibitemOpen
  \bibfield  {author} {\bibinfo {author} {\bibfnamefont {C.~S.}\ \bibnamefont
  {O'Hern}}, \bibinfo {author} {\bibfnamefont {L.~E.}\ \bibnamefont {Silbert}},
  \bibinfo {author} {\bibfnamefont {A.~J.}\ \bibnamefont {Liu}}, \ and\
  \bibinfo {author} {\bibfnamefont {S.~R.}\ \bibnamefont {Nagel}},\ }\href
  {\doibase 10.1103/PhysRevE.68.011306} {\bibfield  {journal} {\bibinfo
  {journal} {Phys. Rev. E}\ }\textbf {\bibinfo {volume} {68}},\ \bibinfo
  {pages} {011306} (\bibinfo {year} {2003})}\BibitemShut {NoStop}%
\bibitem [{\citenamefont {Chen}\ \emph {et~al.}(2008)\citenamefont {Chen},
  \citenamefont {Holbrook}, \citenamefont {Boolchand}, \citenamefont
  {Georgiev}, \citenamefont {Jackson},\ and\ \citenamefont
  {Micoulaut}}]{Chen08}%
  \BibitemOpen
  \bibfield  {author} {\bibinfo {author} {\bibfnamefont {P.}~\bibnamefont
  {Chen}}, \bibinfo {author} {\bibfnamefont {C.}~\bibnamefont {Holbrook}},
  \bibinfo {author} {\bibfnamefont {P.}~\bibnamefont {Boolchand}}, \bibinfo
  {author} {\bibfnamefont {D.~G.}\ \bibnamefont {Georgiev}}, \bibinfo {author}
  {\bibfnamefont {K.~A.}\ \bibnamefont {Jackson}}, \ and\ \bibinfo {author}
  {\bibfnamefont {M.}~\bibnamefont {Micoulaut}},\ }\href {\doibase
  10.1103/PhysRevB.78.224208} {\bibfield  {journal} {\bibinfo  {journal} {Phys.
  Rev. B}\ }\textbf {\bibinfo {volume} {78}},\ \bibinfo {pages} {224208}
  (\bibinfo {year} {2008})}\BibitemShut {NoStop}%
\bibitem [{\citenamefont {Chen}\ \emph {et~al.}(2010)\citenamefont {Chen},
  \citenamefont {Ellenbroek}, \citenamefont {Zhang}, \citenamefont {Chen},
  \citenamefont {Yunker}, \citenamefont {Henkes}, \citenamefont {Brito},
  \citenamefont {Dauchot}, \citenamefont {Van~Saarloos}, \citenamefont {Liu}
  \emph {et~al.}}]{Chen10}%
  \BibitemOpen
  \bibfield  {author} {\bibinfo {author} {\bibfnamefont {K.}~\bibnamefont
  {Chen}}, \bibinfo {author} {\bibfnamefont {W.~G.}\ \bibnamefont
  {Ellenbroek}}, \bibinfo {author} {\bibfnamefont {Z.}~\bibnamefont {Zhang}},
  \bibinfo {author} {\bibfnamefont {D.~T.}\ \bibnamefont {Chen}}, \bibinfo
  {author} {\bibfnamefont {P.~J.}\ \bibnamefont {Yunker}}, \bibinfo {author}
  {\bibfnamefont {S.}~\bibnamefont {Henkes}}, \bibinfo {author} {\bibfnamefont
  {C.}~\bibnamefont {Brito}}, \bibinfo {author} {\bibfnamefont
  {O.}~\bibnamefont {Dauchot}}, \bibinfo {author} {\bibfnamefont
  {W.}~\bibnamefont {Van~Saarloos}}, \bibinfo {author} {\bibfnamefont {A.~J.}\
  \bibnamefont {Liu}},  \emph {et~al.},\ }\href@noop {} {\bibfield  {journal}
  {\bibinfo  {journal} {Physical review letters}\ }\textbf {\bibinfo {volume}
  {105}},\ \bibinfo {pages} {025501} (\bibinfo {year} {2010})}\BibitemShut
  {NoStop}%
\bibitem [{\citenamefont {Ghosh}\ \emph {et~al.}(2010)\citenamefont {Ghosh},
  \citenamefont {Chikkadi}, \citenamefont {Schall}, \citenamefont {Kurchan},\
  and\ \citenamefont {Bonn}}]{Ghosh10}%
  \BibitemOpen
  \bibfield  {author} {\bibinfo {author} {\bibfnamefont {A.}~\bibnamefont
  {Ghosh}}, \bibinfo {author} {\bibfnamefont {V.~K.}\ \bibnamefont {Chikkadi}},
  \bibinfo {author} {\bibfnamefont {P.}~\bibnamefont {Schall}}, \bibinfo
  {author} {\bibfnamefont {J.}~\bibnamefont {Kurchan}}, \ and\ \bibinfo
  {author} {\bibfnamefont {D.}~\bibnamefont {Bonn}},\ }\href@noop {} {\bibfield
   {journal} {\bibinfo  {journal} {Physical review letters}\ }\textbf {\bibinfo
  {volume} {104}},\ \bibinfo {pages} {248305} (\bibinfo {year}
  {2010})}\BibitemShut {NoStop}%
\bibitem [{\citenamefont {Wyart}\ \emph {et~al.}(2005)\citenamefont {Wyart},
  \citenamefont {Silbert}, \citenamefont {Nagel},\ and\ \citenamefont
  {Witten}}]{Wyart05a}%
  \BibitemOpen
  \bibfield  {author} {\bibinfo {author} {\bibfnamefont {M.}~\bibnamefont
  {Wyart}}, \bibinfo {author} {\bibfnamefont {L.~E.}\ \bibnamefont {Silbert}},
  \bibinfo {author} {\bibfnamefont {S.~R.}\ \bibnamefont {Nagel}}, \ and\
  \bibinfo {author} {\bibfnamefont {T.~A.}\ \bibnamefont {Witten}},\
  }\href@noop {} {\bibfield  {journal} {\bibinfo  {journal} {Physical Review
  E}\ }\textbf {\bibinfo {volume} {72}},\ \bibinfo {pages} {051306} (\bibinfo
  {year} {2005})}\BibitemShut {NoStop}%
\bibitem [{\citenamefont {Wyart}(2005)}]{Wyart05b}%
  \BibitemOpen
  \bibfield  {author} {\bibinfo {author} {\bibfnamefont {M.}~\bibnamefont
  {Wyart}},\ }\href@noop {} {\bibfield  {journal} {\bibinfo  {journal} {Annales
  de Phys}\ }\textbf {\bibinfo {volume} {30}},\ \bibinfo {pages} {1} (\bibinfo
  {year} {2005})}\BibitemShut {NoStop}%
\bibitem [{\citenamefont {Xu}\ \emph {et~al.}(2007)\citenamefont {Xu},
  \citenamefont {Wyart}, \citenamefont {Liu},\ and\ \citenamefont
  {Nagel}}]{Xu07}%
  \BibitemOpen
  \bibfield  {author} {\bibinfo {author} {\bibfnamefont {N.}~\bibnamefont
  {Xu}}, \bibinfo {author} {\bibfnamefont {M.}~\bibnamefont {Wyart}}, \bibinfo
  {author} {\bibfnamefont {A.~J.}\ \bibnamefont {Liu}}, \ and\ \bibinfo
  {author} {\bibfnamefont {S.~R.}\ \bibnamefont {Nagel}},\ }\href@noop {}
  {\bibfield  {journal} {\bibinfo  {journal} {Physical review letters}\
  }\textbf {\bibinfo {volume} {98}},\ \bibinfo {pages} {175502} (\bibinfo
  {year} {2007})}\BibitemShut {NoStop}%
\bibitem [{\citenamefont {Brito}\ and\ \citenamefont {Wyart}(2009)}]{Brito09}%
  \BibitemOpen
  \bibfield  {author} {\bibinfo {author} {\bibfnamefont {C.}~\bibnamefont
  {Brito}}\ and\ \bibinfo {author} {\bibfnamefont {M.}~\bibnamefont {Wyart}},\
  }\href {\doibase 10.1063/1.3157261} {\bibfield  {journal} {\bibinfo
  {journal} {The Journal of Chemical Physics}\ }\textbf {\bibinfo {volume}
  {131}},\ \bibinfo {eid} {024504} (\bibinfo {year} {2009})}\BibitemShut
  {NoStop}%
\bibitem [{\citenamefont {de~Souza}\ and\ \citenamefont
  {Harrowell}(2009{\natexlab{b}})}]{Souza09a}%
  \BibitemOpen
  \bibfield  {author} {\bibinfo {author} {\bibfnamefont {V.~K.}\ \bibnamefont
  {de~Souza}}\ and\ \bibinfo {author} {\bibfnamefont {P.}~\bibnamefont
  {Harrowell}},\ }\href {\doibase 10.1103/PhysRevE.80.041503} {\bibfield
  {journal} {\bibinfo  {journal} {Phys. Rev. E}\ }\textbf {\bibinfo {volume}
  {80}},\ \bibinfo {pages} {041503} (\bibinfo {year}
  {2009}{\natexlab{b}})}\BibitemShut {NoStop}%
\bibitem [{\citenamefont {DeGiuli}\ \emph
  {et~al.}(2014{\natexlab{a}})\citenamefont {DeGiuli}, \citenamefont
  {Laversanne-Finot}, \citenamefont {D\"uring}, \citenamefont {Lerner},\ and\
  \citenamefont {Wyart}}]{DeGiuli14}%
  \BibitemOpen
  \bibfield  {author} {\bibinfo {author} {\bibfnamefont {E.}~\bibnamefont
  {DeGiuli}}, \bibinfo {author} {\bibfnamefont {A.}~\bibnamefont
  {Laversanne-Finot}}, \bibinfo {author} {\bibfnamefont {G.~A.}\ \bibnamefont
  {D\"uring}}, \bibinfo {author} {\bibfnamefont {E.}~\bibnamefont {Lerner}}, \
  and\ \bibinfo {author} {\bibfnamefont {M.}~\bibnamefont {Wyart}},\
  }\href@noop {} {\bibfield  {journal} {\bibinfo  {journal} {Soft Matter}\
  }\textbf {\bibinfo {volume} {10}},\ \bibinfo {pages} {5628} (\bibinfo {year}
  {2014}{\natexlab{a}})}\BibitemShut {NoStop}%
\bibitem [{\citenamefont {DeGiuli}\ \emph
  {et~al.}(2014{\natexlab{b}})\citenamefont {DeGiuli}, \citenamefont {Lerner},
  \citenamefont {Brito},\ and\ \citenamefont {Wyart}}]{DeGiuli14b}%
  \BibitemOpen
  \bibfield  {author} {\bibinfo {author} {\bibfnamefont {E.}~\bibnamefont
  {DeGiuli}}, \bibinfo {author} {\bibfnamefont {E.}~\bibnamefont {Lerner}},
  \bibinfo {author} {\bibfnamefont {C.}~\bibnamefont {Brito}}, \ and\ \bibinfo
  {author} {\bibfnamefont {M.}~\bibnamefont {Wyart}},\ }\href@noop {}
  {\bibfield  {journal} {\bibinfo  {journal} {Proceedings of the National
  Academy of Sciences}\ }\textbf {\bibinfo {volume} {111}},\ \bibinfo {pages}
  {17054} (\bibinfo {year} {2014}{\natexlab{b}})}\BibitemShut {NoStop}%
\bibitem [{\citenamefont {DeGiuli}\ \emph {et~al.}(2015)\citenamefont
  {DeGiuli}, \citenamefont {Lerner},\ and\ \citenamefont {Wyart}}]{Degiuli15}%
  \BibitemOpen
  \bibfield  {author} {\bibinfo {author} {\bibfnamefont {E.}~\bibnamefont
  {DeGiuli}}, \bibinfo {author} {\bibfnamefont {E.}~\bibnamefont {Lerner}}, \
  and\ \bibinfo {author} {\bibfnamefont {M.}~\bibnamefont {Wyart}},\
  }\href@noop {} {\bibfield  {journal} {\bibinfo  {journal} {The Journal of
  chemical physics}\ }\textbf {\bibinfo {volume} {142}},\ \bibinfo {pages}
  {164503} (\bibinfo {year} {2015})}\BibitemShut {NoStop}%
\bibitem [{\citenamefont {{Franz}}\ \emph {et~al.}(2015)\citenamefont
  {{Franz}}, \citenamefont {{Parisi}}, \citenamefont {{Urbani}},\ and\
  \citenamefont {{Zamponi}}}]{Franz15b}%
  \BibitemOpen
  \bibfield  {author} {\bibinfo {author} {\bibfnamefont {S.}~\bibnamefont
  {{Franz}}}, \bibinfo {author} {\bibfnamefont {G.}~\bibnamefont {{Parisi}}},
  \bibinfo {author} {\bibfnamefont {P.}~\bibnamefont {{Urbani}}}, \ and\
  \bibinfo {author} {\bibfnamefont {F.}~\bibnamefont {{Zamponi}}},\ }\href@noop
  {} {\bibfield  {journal} {\bibinfo  {journal} {ArXiv e-prints}\ } (\bibinfo
  {year} {2015})},\ \Eprint {http://arxiv.org/abs/1506.01997} {arXiv:1506.01997
  [cond-mat.dis-nn]} \BibitemShut {NoStop}%
\bibitem [{\citenamefont {Yan}\ \emph {et~al.}(2013)\citenamefont {Yan},
  \citenamefont {D{\"u}ring},\ and\ \citenamefont {Wyart}}]{Yan13}%
  \BibitemOpen
  \bibfield  {author} {\bibinfo {author} {\bibfnamefont {L.}~\bibnamefont
  {Yan}}, \bibinfo {author} {\bibfnamefont {G.}~\bibnamefont {D{\"u}ring}}, \
  and\ \bibinfo {author} {\bibfnamefont {M.}~\bibnamefont {Wyart}},\
  }\href@noop {} {\bibfield  {journal} {\bibinfo  {journal} {Proceedings of the
  National Academy of Sciences}\ }\textbf {\bibinfo {volume} {110}},\ \bibinfo
  {pages} {6307} (\bibinfo {year} {2013})}\BibitemShut {NoStop}%
\bibitem [{\citenamefont {Yan}\ and\ \citenamefont {Wyart}(2014)}]{Yan14}%
  \BibitemOpen
  \bibfield  {author} {\bibinfo {author} {\bibfnamefont {L.}~\bibnamefont
  {Yan}}\ and\ \bibinfo {author} {\bibfnamefont {M.}~\bibnamefont {Wyart}},\
  }\href {\doibase 10.1103/PhysRevLett.113.215504} {\bibfield  {journal}
  {\bibinfo  {journal} {Phys. Rev. Lett.}\ }\textbf {\bibinfo {volume} {113}},\
  \bibinfo {pages} {215504} (\bibinfo {year} {2014})}\BibitemShut {NoStop}%
\bibitem [{\citenamefont {Ellenbroek}\ \emph {et~al.}(2015)\citenamefont
  {Ellenbroek}, \citenamefont {Hagh}, \citenamefont {Kumar}, \citenamefont
  {Thorpe},\ and\ \citenamefont {van Hecke}}]{Ellenbroek15}%
  \BibitemOpen
  \bibfield  {author} {\bibinfo {author} {\bibfnamefont {W.~G.}\ \bibnamefont
  {Ellenbroek}}, \bibinfo {author} {\bibfnamefont {V.~F.}\ \bibnamefont
  {Hagh}}, \bibinfo {author} {\bibfnamefont {A.}~\bibnamefont {Kumar}},
  \bibinfo {author} {\bibfnamefont {M.~F.}\ \bibnamefont {Thorpe}}, \ and\
  \bibinfo {author} {\bibfnamefont {M.}~\bibnamefont {van Hecke}},\ }\href
  {\doibase 10.1103/PhysRevLett.114.135501} {\bibfield  {journal} {\bibinfo
  {journal} {Phys. Rev. Lett.}\ }\textbf {\bibinfo {volume} {114}},\ \bibinfo
  {pages} {135501} (\bibinfo {year} {2015})}\BibitemShut {NoStop}%
\bibitem [{\citenamefont {Moukarzel}(2013)}]{Moukarzel13}%
  \BibitemOpen
  \bibfield  {author} {\bibinfo {author} {\bibfnamefont {C.~F.}\ \bibnamefont
  {Moukarzel}},\ }\href {\doibase 10.1103/PhysRevE.88.062121} {\bibfield
  {journal} {\bibinfo  {journal} {Phys. Rev. E}\ }\textbf {\bibinfo {volume}
  {88}},\ \bibinfo {pages} {062121} (\bibinfo {year} {2013})}\BibitemShut
  {NoStop}%
\bibitem [{\citenamefont {Jacobs}\ and\ \citenamefont
  {Thorpe}(1998)}]{Jacobs98}%
  \BibitemOpen
  \bibfield  {author} {\bibinfo {author} {\bibfnamefont {D.~J.}\ \bibnamefont
  {Jacobs}}\ and\ \bibinfo {author} {\bibfnamefont {M.~F.}\ \bibnamefont
  {Thorpe}},\ }\href {\doibase 10.1103/PhysRevLett.80.5451} {\bibfield
  {journal} {\bibinfo  {journal} {Phys. Rev. Lett.}\ }\textbf {\bibinfo
  {volume} {80}},\ \bibinfo {pages} {5451} (\bibinfo {year}
  {1998})}\BibitemShut {NoStop}%
\bibitem [{\citenamefont {Duxbury}\ \emph {et~al.}(1999)\citenamefont
  {Duxbury}, \citenamefont {Jacobs}, \citenamefont {Thorpe},\ and\
  \citenamefont {Moukarzel}}]{Duxbury99}%
  \BibitemOpen
  \bibfield  {author} {\bibinfo {author} {\bibfnamefont {P.~M.}\ \bibnamefont
  {Duxbury}}, \bibinfo {author} {\bibfnamefont {D.~J.}\ \bibnamefont {Jacobs}},
  \bibinfo {author} {\bibfnamefont {M.~F.}\ \bibnamefont {Thorpe}}, \ and\
  \bibinfo {author} {\bibfnamefont {C.}~\bibnamefont {Moukarzel}},\ }\href
  {\doibase 10.1103/PhysRevE.59.2084} {\bibfield  {journal} {\bibinfo
  {journal} {Phys. Rev. E}\ }\textbf {\bibinfo {volume} {59}},\ \bibinfo
  {pages} {2084} (\bibinfo {year} {1999})}\BibitemShut {NoStop}%
\bibitem [{\citenamefont {Feng}\ and\ \citenamefont {Sen}(1984)}]{Feng84}%
  \BibitemOpen
  \bibfield  {author} {\bibinfo {author} {\bibfnamefont {S.}~\bibnamefont
  {Feng}}\ and\ \bibinfo {author} {\bibfnamefont {P.~N.}\ \bibnamefont {Sen}},\
  }\href {\doibase 10.1103/PhysRevLett.52.216} {\bibfield  {journal} {\bibinfo
  {journal} {Phys. Rev. Lett.}\ }\textbf {\bibinfo {volume} {52}},\ \bibinfo
  {pages} {216} (\bibinfo {year} {1984})}\BibitemShut {NoStop}%
\bibitem [{\citenamefont {Jacobs}\ and\ \citenamefont
  {Thorpe}(1995)}]{Jacobs95}%
  \BibitemOpen
  \bibfield  {author} {\bibinfo {author} {\bibfnamefont {D.~J.}\ \bibnamefont
  {Jacobs}}\ and\ \bibinfo {author} {\bibfnamefont {M.~F.}\ \bibnamefont
  {Thorpe}},\ }\href {\doibase 10.1103/PhysRevLett.75.4051} {\bibfield
  {journal} {\bibinfo  {journal} {Phys. Rev. Lett.}\ }\textbf {\bibinfo
  {volume} {75}},\ \bibinfo {pages} {4051} (\bibinfo {year}
  {1995})}\BibitemShut {NoStop}%
\bibitem [{\citenamefont {Thorpe}\ \emph {et~al.}(2000)\citenamefont {Thorpe},
  \citenamefont {Jacobs}, \citenamefont {Chubynsky},\ and\ \citenamefont
  {Phillips}}]{Thorpe00}%
  \BibitemOpen
  \bibfield  {author} {\bibinfo {author} {\bibfnamefont {M.}~\bibnamefont
  {Thorpe}}, \bibinfo {author} {\bibfnamefont {D.}~\bibnamefont {Jacobs}},
  \bibinfo {author} {\bibfnamefont {M.}~\bibnamefont {Chubynsky}}, \ and\
  \bibinfo {author} {\bibfnamefont {J.}~\bibnamefont {Phillips}},\ }\href
  {\doibase http://dx.doi.org/10.1016/S0022-3093(99)00856-X} {\bibfield
  {journal} {\bibinfo  {journal} {Journal of Non-Crystalline Solids}\ }\textbf
  {\bibinfo {volume} {266-269, Part 2}},\ \bibinfo {pages} {859 } (\bibinfo
  {year} {2000})}\BibitemShut {NoStop}%
\bibitem [{\citenamefont {Chubynsky}\ \emph {et~al.}(2006)\citenamefont
  {Chubynsky}, \citenamefont {Bri\`ere},\ and\ \citenamefont
  {Mousseau}}]{Chubynsky06}%
  \BibitemOpen
  \bibfield  {author} {\bibinfo {author} {\bibfnamefont {M.~V.}\ \bibnamefont
  {Chubynsky}}, \bibinfo {author} {\bibfnamefont {M.-A.}\ \bibnamefont
  {Bri\`ere}}, \ and\ \bibinfo {author} {\bibfnamefont {N.}~\bibnamefont
  {Mousseau}},\ }\href {\doibase 10.1103/PhysRevE.74.016116} {\bibfield
  {journal} {\bibinfo  {journal} {Phys. Rev. E}\ }\textbf {\bibinfo {volume}
  {74}},\ \bibinfo {pages} {016116} (\bibinfo {year} {2006})}\BibitemShut
  {NoStop}%
\bibitem [{\citenamefont {Bri\`ere}\ \emph {et~al.}(2007)\citenamefont
  {Bri\`ere}, \citenamefont {Chubynsky},\ and\ \citenamefont
  {Mousseau}}]{Briere07}%
  \BibitemOpen
  \bibfield  {author} {\bibinfo {author} {\bibfnamefont {M.-A.}\ \bibnamefont
  {Bri\`ere}}, \bibinfo {author} {\bibfnamefont {M.~V.}\ \bibnamefont
  {Chubynsky}}, \ and\ \bibinfo {author} {\bibfnamefont {N.}~\bibnamefont
  {Mousseau}},\ }\href {\doibase 10.1103/PhysRevE.75.056108} {\bibfield
  {journal} {\bibinfo  {journal} {Phys. Rev. E}\ }\textbf {\bibinfo {volume}
  {75}},\ \bibinfo {pages} {056108} (\bibinfo {year} {2007})}\BibitemShut
  {NoStop}%
\bibitem [{\citenamefont {Micoulaut}\ and\ \citenamefont
  {Phillips}(2003)}]{Micoulaut03}%
  \BibitemOpen
  \bibfield  {author} {\bibinfo {author} {\bibfnamefont {M.}~\bibnamefont
  {Micoulaut}}\ and\ \bibinfo {author} {\bibfnamefont {J.~C.}\ \bibnamefont
  {Phillips}},\ }\href {\doibase 10.1103/PhysRevB.67.104204} {\bibfield
  {journal} {\bibinfo  {journal} {Phys. Rev. B}\ }\textbf {\bibinfo {volume}
  {67}},\ \bibinfo {pages} {104204} (\bibinfo {year} {2003})}\BibitemShut
  {NoStop}%
\bibitem [{Note1()}]{Note1}%
  \BibitemOpen
  \bibinfo {note} {We have tested the validity of the linear approximation: the
  energy difference from the steepest decent results keeps below 3\% for
  $\epsilon <0.02$.}\BibitemShut {Stop}%
\bibitem [{\citenamefont {Barr\'e}(2009)}]{Barre09}%
  \BibitemOpen
  \bibfield  {author} {\bibinfo {author} {\bibfnamefont {J.}~\bibnamefont
  {Barr\'e}},\ }\href {\doibase 10.1103/PhysRevE.80.061108} {\bibfield
  {journal} {\bibinfo  {journal} {Phys. Rev. E}\ }\textbf {\bibinfo {volume}
  {80}},\ \bibinfo {pages} {061108} (\bibinfo {year} {2009})}\BibitemShut
  {NoStop}%
\bibitem [{\citenamefont {Jacobs}\ and\ \citenamefont
  {Hendrickson}(1997)}]{Jacobs97}%
  \BibitemOpen
  \bibfield  {author} {\bibinfo {author} {\bibfnamefont {D.~J.}\ \bibnamefont
  {Jacobs}}\ and\ \bibinfo {author} {\bibfnamefont {B.}~\bibnamefont
  {Hendrickson}},\ }\href {\doibase http://dx.doi.org/10.1006/jcph.1997.5809}
  {\bibfield  {journal} {\bibinfo  {journal} {Journal of Computational
  Physics}\ }\textbf {\bibinfo {volume} {137}},\ \bibinfo {pages} {346 }
  (\bibinfo {year} {1997})}\BibitemShut {NoStop}%
\bibitem [{\citenamefont {Grigera}\ and\ \citenamefont
  {Parisi}(2001)}]{Grigera01a}%
  \BibitemOpen
  \bibfield  {author} {\bibinfo {author} {\bibfnamefont {T.~S.}\ \bibnamefont
  {Grigera}}\ and\ \bibinfo {author} {\bibfnamefont {G.}~\bibnamefont
  {Parisi}},\ }\href {\doibase 10.1103/PhysRevE.63.045102} {\bibfield
  {journal} {\bibinfo  {journal} {Phys. Rev. E}\ }\textbf {\bibinfo {volume}
  {63}},\ \bibinfo {pages} {045102} (\bibinfo {year} {2001})}\BibitemShut
  {NoStop}%
\bibitem [{\citenamefont {Lindemann}(1910)}]{Lindemann10}%
  \BibitemOpen
  \bibfield  {author} {\bibinfo {author} {\bibfnamefont {F.}~\bibnamefont
  {Lindemann}},\ }\href@noop {} {\bibfield  {journal} {\bibinfo  {journal} {Z.
  Phys.}\ }\textbf {\bibinfo {volume} {11}},\ \bibinfo {pages} {609} (\bibinfo
  {year} {1910})}\BibitemShut {NoStop}%
\bibitem [{\citenamefont {Nelson}(2002)}]{Nelson02}%
  \BibitemOpen
  \bibfield  {author} {\bibinfo {author} {\bibfnamefont {D.~R.}\ \bibnamefont
  {Nelson}},\ }\href@noop {} {\emph {\bibinfo {title} {Defects and geometry in
  condensed matter physics}}}\ (\bibinfo  {publisher} {Cambridge University
  Press},\ \bibinfo {year} {2002})\BibitemShut {NoStop}%
\bibitem [{\citenamefont {Derrida}(1981)}]{Derrida81}%
  \BibitemOpen
  \bibfield  {author} {\bibinfo {author} {\bibfnamefont {B.}~\bibnamefont
  {Derrida}},\ }\href {\doibase 10.1103/PhysRevB.24.2613} {\bibfield  {journal}
  {\bibinfo  {journal} {Phys. Rev. B}\ }\textbf {\bibinfo {volume} {24}},\
  \bibinfo {pages} {2613} (\bibinfo {year} {1981})}\BibitemShut {NoStop}%
\bibitem [{\citenamefont {M\'ezard}(2009)}]{Mezard09}%
  \BibitemOpen
  \bibfield  {author} {\bibinfo {author} {\bibfnamefont {M.~a.}\ \bibnamefont
  {M\'ezard}},\ }\href@noop {} {\emph {\bibinfo {title} {Information, Physics
  and Computation}}}\ (\bibinfo  {publisher} {Oxford University press},\
  \bibinfo {year} {2009})\BibitemShut {NoStop}%
\bibitem [{\citenamefont {{Wyart}}\ \emph {et~al.}(2008)\citenamefont
  {{Wyart}}, \citenamefont {{Liang}}, \citenamefont {{Kabla}},\ and\
  \citenamefont {{Mahadevan}}}]{Wyart08}%
  \BibitemOpen
  \bibfield  {author} {\bibinfo {author} {\bibfnamefont {M.}~\bibnamefont
  {{Wyart}}}, \bibinfo {author} {\bibfnamefont {H.}~\bibnamefont {{Liang}}},
  \bibinfo {author} {\bibfnamefont {A.}~\bibnamefont {{Kabla}}}, \ and\
  \bibinfo {author} {\bibfnamefont {L.}~\bibnamefont {{Mahadevan}}},\
  }\href@noop {} {\bibfield  {journal} {\bibinfo  {journal} {Phys.\ Rev.\
  Lett.}\ }\textbf {\bibinfo {volume} {101}},\ \bibinfo {pages} {215501}
  (\bibinfo {year} {2008})}\BibitemShut {NoStop}%
\bibitem [{\citenamefont {D{\"u}ring}\ \emph {et~al.}(2013)\citenamefont
  {D{\"u}ring}, \citenamefont {Lerner},\ and\ \citenamefont
  {Wyart}}]{During13}%
  \BibitemOpen
  \bibfield  {author} {\bibinfo {author} {\bibfnamefont {G.}~\bibnamefont
  {D{\"u}ring}}, \bibinfo {author} {\bibfnamefont {E.}~\bibnamefont {Lerner}},
  \ and\ \bibinfo {author} {\bibfnamefont {M.}~\bibnamefont {Wyart}},\
  }\href@noop {} {\bibfield  {journal} {\bibinfo  {journal} {Soft Matter}\
  }\textbf {\bibinfo {volume} {9}},\ \bibinfo {pages} {146} (\bibinfo {year}
  {2013})}\BibitemShut {NoStop}%
\bibitem [{\citenamefont {Boolchand}\ \emph {et~al.}(2005)\citenamefont
  {Boolchand}, \citenamefont {Lucovsky}, \citenamefont {Phillips},\ and\
  \citenamefont {Thorpe}}]{Boolchand05}%
  \BibitemOpen
  \bibfield  {author} {\bibinfo {author} {\bibfnamefont {P.}~\bibnamefont
  {Boolchand}}, \bibinfo {author} {\bibfnamefont {G.}~\bibnamefont {Lucovsky}},
  \bibinfo {author} {\bibfnamefont {J.~C.}\ \bibnamefont {Phillips}}, \ and\
  \bibinfo {author} {\bibfnamefont {M.~F.}\ \bibnamefont {Thorpe}},\ }\href
  {\doibase 10.1080/14786430500256425} {\bibfield  {journal} {\bibinfo
  {journal} {Philosophical Magazine}\ }\textbf {\bibinfo {volume} {85}},\
  \bibinfo {pages} {3823} (\bibinfo {year} {2005})}\BibitemShut {NoStop}%
\bibitem [{\citenamefont {Calladine}(1978)}]{Calladine78}%
  \BibitemOpen
  \bibfield  {author} {\bibinfo {author} {\bibfnamefont {C.}~\bibnamefont
  {Calladine}},\ }\href {\doibase 10.1016/0020-7683(78)90052-5} {\bibfield
  {journal} {\bibinfo  {journal} {International Journal of Solids and
  Structures}\ }\textbf {\bibinfo {volume} {14}},\ \bibinfo {pages} {161 }
  (\bibinfo {year} {1978})}\BibitemShut {NoStop}%
\bibitem [{\citenamefont {Alexander}\ and\ \citenamefont
  {Orbach}(1982)}]{Alexander82}%
  \BibitemOpen
  \bibfield  {author} {\bibinfo {author} {\bibfnamefont {S.}~\bibnamefont
  {Alexander}}\ and\ \bibinfo {author} {\bibfnamefont {R.}~\bibnamefont
  {Orbach}},\ }\href {\doibase 10.1051/jphyslet:019820043017062500} {\bibfield
  {journal} {\bibinfo  {journal} {J. Phys. (Paris) Lett.}\ }\textbf {\bibinfo
  {volume} {43}},\ \bibinfo {pages} {625} (\bibinfo {year} {1982})}\BibitemShut
  {NoStop}%
\bibitem [{\citenamefont {Nakayama}\ \emph {et~al.}(1994)\citenamefont
  {Nakayama}, \citenamefont {Yakubo},\ and\ \citenamefont
  {Orbach}}]{Nakayama94}%
  \BibitemOpen
  \bibfield  {author} {\bibinfo {author} {\bibfnamefont {T.}~\bibnamefont
  {Nakayama}}, \bibinfo {author} {\bibfnamefont {K.}~\bibnamefont {Yakubo}}, \
  and\ \bibinfo {author} {\bibfnamefont {R.~L.}\ \bibnamefont {Orbach}},\
  }\href {\doibase 10.1103/RevModPhys.66.381} {\bibfield  {journal} {\bibinfo
  {journal} {Rev. Mod. Phys.}\ }\textbf {\bibinfo {volume} {66}},\ \bibinfo
  {pages} {381} (\bibinfo {year} {1994})}\BibitemShut {NoStop}%
\bibitem [{\citenamefont {Feng}(1985)}]{Feng85}%
  \BibitemOpen
  \bibfield  {author} {\bibinfo {author} {\bibfnamefont {S.}~\bibnamefont
  {Feng}},\ }\href {\doibase 10.1103/PhysRevB.32.5793} {\bibfield  {journal}
  {\bibinfo  {journal} {Phys. Rev. B}\ }\textbf {\bibinfo {volume} {32}},\
  \bibinfo {pages} {5793} (\bibinfo {year} {1985})}\BibitemShut {NoStop}%
\bibitem [{\citenamefont {Wyart}(2010)}]{Wyart10}%
  \BibitemOpen
  \bibfield  {author} {\bibinfo {author} {\bibfnamefont {M.}~\bibnamefont
  {Wyart}},\ }\href@noop {} {\bibfield  {journal} {\bibinfo  {journal} {Phys.
  Rev. Lett.}\ }\textbf {\bibinfo {volume} {104}},\ \bibinfo {pages} {095901}
  (\bibinfo {year} {2010})}\BibitemShut {NoStop}%
\end{thebibliography}%

\end{document}